\documentclass[aps,prd,amsmath,amssymb,floatfix,nofootinbib,superscriptaddress,longbibliography,11pt]{revtex4-2}
\usepackage{amsmath}
\usepackage{tikz}
\usetikzlibrary{arrows}
\usetikzlibrary{decorations.pathmorphing}
\usepackage{graphicx} 
\usepackage{amssymb}
\usepackage{hyperref}
\usepackage{subfigure}

\usepackage[english]{babel}
\usepackage[utf8]{inputenc}
\usepackage[T1]{fontenc}

\usepackage{bbm}
\usepackage{CJKutf8}
\usepackage{physics}
\usepackage{subfigure}
\usepackage{mathrsfs}
\usepackage{bm}
\usepackage{enumerate}
\usepackage{graphicx}

\allowdisplaybreaks[4]

\newcounter{jvcc}
\newcounter{amgg}
\newcounter{jz}

\renewcommand{\Im}{\mathrm{Im}}

\newcommand{\wSigma}{\widetilde \Sigma}
\renewcommand{\hat}{\widehat}
\renewcommand{\tilde}{\widetilde}
\renewcommand{\bar}{\overline}

\newcommand{\del}{\partial}
\newcommand{\edoc}{\end{document}}
\newcommand{\eref}[1]{(\ref{#1})}
\newcommand{\nn}{\nonumber}

\renewcommand{\i}{\mathrm{i}}

\newcommand{\id}{\mathbbm{1}}

\newcommand{\sC}{\mathcal{C}}

\newcommand{\sgn}{{\rm sgn}}
\newcommand{\oG}{\overline{G}}
\newcommand{\oT}{{\overline{T} }}
\newcommand{\be}{\begin{eqnarray}}
	\newcommand{\bea}{\begin{eqnarray}}
		\newcommand{\eea}{\end{eqnarray}}
	\newcommand{\beq}{\begin{equation}}
		\newcommand{\ee}{\end{eqnarray}}
	\newcommand{\eeq}{\end{equation}}

\newcounter{amg}

\begin{document}
\title{The Lyapunov exponent as a signature of dissipative \\  many-body quantum chaos}
\author{Antonio  M. Garc\'\i a-Garc\'\i a}
\email{amgg@sjtu.edu.cn}
\affiliation{Shanghai Center for Complex Physics,
	School of Physics and Astronomy, Shanghai Jiao Tong
	University, Shanghai 200240, China}
\author{Jacobus J. M. Verbaarschot}
\email{jacobus.verbaarschot@stonybrook.edu}
\affiliation{Center for Nuclear Theory and Department of Physics and Astronomy, Stony Brook University, Stony Brook, New York 11794, USA}
\author{\begin{CJK*}{UTF8}{gbsn}
		Jie-ping Zheng (郑杰平)
\end{CJK*}}
\email{jpzheng@sjtu.edu. cn}
\affiliation{Shanghai Center for Complex Physics,
	School of Physics and Astronomy, Shanghai Jiao Tong
	University, Shanghai 200240, China}

\begin{abstract}
	\vspace{0.5cm}
        A distinct feature of Hermitian quantum chaotic dynamics is the exponential increase of certain out-of-time-order-correlation
        (OTOC) functions around the Ehrenfest time with a rate given by a Lyapunov exponent.
        Physically, the OTOCs describe
        the growth of quantum uncertainty
       that crucially depends on the nature of the quantum motion.
        Here, we employ the OTOC in order to provide a precise definition
        of dissipative quantum chaos. For this purpose,
        we compute analytically the Lyapunov exponent for the vectorized
        formulation of the large $q$-limit of a $q$-body Sachdev-Ye-Kitaev model
        coupled to a Markovian bath. These analytic results are confirmed by an explicit numerical calculation of the Lyapunov exponent for several values of $q \geq 4$ based on the solutions of the Schwinger-Dyson and Bethe-Salpeter equations.
        We show that the Lyapunov exponent decreases monotonically
        as the coupling to the bath increases and eventually becomes negative at a critical value of the coupling signaling a transition to a dynamics which is no longer quantum chaotic.
        Therefore, a positive Lyapunov exponent is a defining feature of dissipative many-body
        quantum chaos. The observation of the breaking of the exponential growth for sufficiently strong coupling suggests that dissipative quantum chaos may require in certain cases a sufficiently weak coupling to the environment.    
\end{abstract}

\maketitle

\newpage

	\tableofcontents
\section{Introduction}
Research on many-body quantum chaos is receiving a lot of attention because its strikingly universal features  allow a qualitative description of the quantum time evolution that does not depend on the details of the Hamiltonian. These universal features leave fingerprints in the quantum dynamics at different time scales.

One of these time scales is the so called  Ehrenfest time,
or scrambling \cite{sekino2008,bagrets2017} time, $t_E$, characterized by the time scale below which the quantum dynamics largely follows the classical motion.
For $t \gtrsim t_E$, deviations from the classical dynamics due to the increase of quantum uncertainty develop rapidly.
For semiclassical quantum chaotic systems, this is reflected in the exponential growth of certain out-of-time correlation (OTOC) functions
\cite{larkin1969,berman1978} with a rate controlled by the classical Lyapunov exponent describing
the exponential divergence of classical trajectories. By contrast, for integrable systems, the growth of the OTOC is generically slower than exponential. This makes
the calculation of the OTOC a powerful tool to characterize the nature of the quantum dynamics around the relatively short time scale of the Ehrenfest time.
 
For much longer times, of the order of the Heisenberg time, the inverse of the mean level spacing, quantum chaotic dynamics is well characterized by level statistics. According to the celebrated Bohigas-Giannoni-Schmit (BGS) conjecture \cite{bohigas1984}, level statistics of quantum chaotic systems agrees  with the predictions of random matrix theory \cite{wigner1951,dyson1962a,dyson1962b,dyson1962c,dyson1962d,dyson1972,mehta2004}. The study of spectral correlations has been by far the most used tool to study quantum chaos because it is relatively easy to obtain the spectrum numerically and compare it with the compact analytic results provided by random matrix theory \cite{mehta2004}.  

However, in the early days of the development of quantum chaos theory, the OTOC, and the Lyapunov exponent, were computed analytically, or semi-analytically, for simple systems such as a kicked rotor \cite{berman1978} or a non-interacting particle in a random potential \cite{larkin1969}. For many-body quantum chaotic systems, it was considered quite challenging to compute the OTOC either numerically or analytically. 

Recently, the situation changed drastically after two major results: the proposal that in quantum chaotic systems the Lyapunov exponent is subject to a
universal bound \cite{maldacena2015} which is conjectured to be saturated in field theories with a gravity dual, and the observation \cite{kitaev2015} that,
in the low temperature limit, the bound is saturated in the now called Sachdev-Ye-Kitaev (SYK) model \cite{french1970,bohigas1971,french1971,bohigas1971a,benet2001,sachdev1993,sachdev2010,kitaev2015,maldacena2016}, $N$ zero-dimensional Majorana fermions with random
$q$-body interactions of infinite range in Fock space. Moreover, recent developments in numerical techniques have made it possible to largely confirm these theoretical predictions \cite{Kobrin:2020xms,Garcia-Garcia:2023jlu}.

Therefore, the SYK model is quantum chaotic at the relatively short time scale of the
Ehrenfest time. It is indeed quantum chaotic at all scales as spectral correlations \cite{garcia2016} are well described by random matrix theory.

All the above results apply only to Hermitian systems. However, 
the dynamics of a quantum chaotic system coupled to an environment, so that the system
Hamiltonian is not Hermitian, has received a recent boost of attention, especially for systems that are PT symmetric \cite{bender1998}, in a broad variety of problems from quantum optics \cite{wiersig2014,gu2020} and condensed matter \cite{kawabata2018,kawabata2017,Roccati2024}, to cold atom \cite{li2019}, and high energy physics \cite{garcia2021}, see Ref.~\cite{ashida2020} for a comprehensive review. 
 
However, the meaning and characterization of quantum chaos in these systems are much less developed. Although the BGS conjecture has effectively been extended to non-Hermitian systems \cite{grobe1988,fyodorov1997}, and there have been several successful comparisons between quantum dissipative systems \cite{sa2019,sa2020,rubio2022,garcia2022d,li2021a,altland2021symmetry,costa2023} with the corresponding non-Hermitian random matrix ensembles \cite{ueda2019,ginibre1965,garcia2002a}, it still remains poorly understood the degree of universality \cite{garcia2023e} of the conjecture in non-Hermitian systems and, more generally, the fate of quantum chaotic effects in the presence of an environment. For instance, one would expect that a strong coupling to the bath would erase, partially or completely, quantum features
but it is unclear how this would precisely be reflected in observables like the OTOC.

Here, we address this problem by studying the OTOC of an SYK model coupled to a Markovian bath \cite{sa2022,kulkarni2022,garcia2022e} in the framework of the so called Lindblad formalism ~\cite{belavin1969,lindblad1976,gorini1976,breuer2002,manzano2020}. 
	
We aim to answer the following questions:
does the OTOC still grow exponentially around the Ehrenfest time in the presence of an environment? If this applies, it could be used as a sharp definition of dissipative quantum chaos. Does a quantum-classical transition occur for a sufficiently strong coupling to the environment? If so, this will constrain the definition of dissipative quantum chaos to the quantum region where the Lyapunov exponent is positive, no matter whether spectral correlations are described by non-Hermitian random matrix theory \cite{ginibre1965}. 

The role of decoherence in the OTOC 
  \cite{bergamasco2023quantum,yoshida2019disentangling,tuziemski2019out,weinstein2023scrambling}
and whether chaotic and integrable motion can be distinguishable in a quantum dissipative system \cite{zanardi2021information} have already been investigated in the literature but, as far as we
know, the question stated above has not yet been fully settled.

Other studies of OTOCs in  non-Hermitian quantum chaotic systems, limited to single particle kicked rotors \cite{huang2020}, two level systems \cite{syzranov2018out}, the Ising model \cite{zhai2020a}, or considering non-Hermitian operators \cite{turiaci2019a} in the OTOC, altogether avoided  addressing the question of the definition of dissipative many-body quantum chaos that motivates our research.
Likewise, different aspects of the dynamics of non-Hermitian SYK's \cite{garcia2022c,Cai2022,schuster:2022bot,bhattacharya2022,bhattacharjee2023,garcia2022e,pengfei2021,pengfei2021c,bhattacharjee2023a,liu2023,srivatsa2023,pengfei2023,liu2024} that have confirmed agreement with the RMT predictions, have also been the subject of recent research, but their emphasis was not on the description of early signatures of dissipative quantum chaos. 

The Lyapunov exponent has been computed \cite{chen2017,maldacena2018,caceres2021,nosaka2021,nosaka2022} in several settings involving two coupled SYK models that, with a certain choice of operators,
could be understood \cite{caldeira1981} as a single site SYK model coupled to a bath modeled by the other
SYK model. This relation has been  explicitly exploited in Ref.~\cite{chen2017} to
show that, in agreement with our results, if the SYK model acting as a bath has a number of Majoranas parametrically larger than the SYK acting as a system, the Lyapunov exponent vanishes if the coupling to the bath is sufficiently strong.   
 
The manuscript is organized as follows: 
we start in section \ref{sec:model} with the introduction of the Lindblad SYK model and its main features.  All calculations are performed for the Keldysh formulation of 
the vectorized form of this model, which
is also discussed in this section.
In section \ref{sec:otoc}, we first carry out the calculation of the OTOC from the numerical solution
of the  Schwinger-Dyson (SD) equations for $q \geq 4$. The resulting Lyapunov exponent
decreases monotonously with the coupling to the bath and eventually vanishes, indicating the transition to a non quantum-chaotic time evolution. Section \ref{sec:otoclargeq} is devoted to an analytical calculation of the Lyapunov exponent in the large $q$ limit that is in full agreement with the numerical findings of the previous section.  
In view of these results, in section \ref{sec:conclusions} we discuss both a possible definition of quantum chaos in dissipative systems and the universality of our results. 
In the appendix, we cover technical aspects of the calculation of the OTOC such as the derivation of both the Schwinger-Dyson and the Bethe-Salpeter equations, the latter also termed kernel equation, necessary to compute the Lyapunov exponent analytically or numerically.   

\section{The dissipative SYK model and the calculation of Green's functions}\label{sec:model}
We study a single-site SYK model coupled to a Markovian
environment described by
the Lindblad formalism \cite{kulkarni2022,sa2022} in which the bath is characterized by jump operators $L_i$ that depend on the single-site SYK Majoranas.
The evolution of the density matrix
of this system is dictated by: 
\begin{equation}\begin{aligned} 
& \frac{d\rho}{dt} = \mathcal{L}, \qquad \mathcal{L} = -i[H,\rho] + \sum_i L_i \rho L_i^{\dagger} - \frac{1}{2}\{L_i^{\dagger}L_i,\rho\} ,
\label{eq:lindblad}
\end{aligned}\end{equation}
where $\mathcal{L}$ is termed the Liouvillian.
The  Hamiltonian $H$
of the SYK model is
given  by,
\begin{equation}\begin{aligned} \label{eq:singleSYK}
    H = i^{q/2}\sum_{i_1<i_2<\cdots <i_q} J_{i_1
   \cdots i_q}\psi_{i_1} \cdots \psi_{i_q} 
\end{aligned}\end{equation}
with  $J_{i_1
	\cdots i_q}$ Gaussian random couplings 
\begin{equation}\begin{aligned} 
\langle J_{i_1i_2\cdots i_q} \rangle=0, \qquad   \langle J_{ijkl}^2 \rangle=\frac{(q-1)!J^2}{N^{q-1}}
\end{aligned}\end{equation}
and $\{\psi_i,\psi_j\}=\delta_{ij}$.
We consider jump operators,
$
L_i=\sqrt{\mu}\psi_i$, linear in the Majorana fermions.  
Since $\psi_i^2 =1/2$, the Liouvillian simplifies to
\begin{equation}\begin{aligned} 
    \qquad \mathcal{L}
    = -i[H,\rho] + \mu\sum_i \psi_i \rho \psi_i - \frac{N\mu}{2}\rho .
\end{aligned}\end{equation}
It is clear that an equilibrium solution of the Lindblad equation is given by
the identity matrix
\be
\rho = \sum_k |k\rangle \langle k|,
\ee
which, for an equilibrium system, is the density matrix at infinite temperature. Physically, this means that the effect of the coupling to a bath induces an out of equilibrium dynamics resulting in an increase of temperature until the final steady state is reached at infinite temperature.  

The study of the dynamics of this system requires \cite{kamenevbook} the vectorization of the Liouvillian, resulting in a doubling of degrees of freedom, in order to set up the Keldysh path integral \cite{sieberer2016}.
This doubling is a direct consequence of the fact that the density matrix in Eq.~(\ref{eq:lindblad})
is an operator that evolves from the right
with the adjoint of the left time evolution, which in the doubled space can be interpreted as backward time evolution. The Liouvillian is usually referred to as a super-operator as it generates the dynamics of another operator, the density matrix. 
We note that in this formulation, the infinite
temperature thermal density matrix $\rho(\infty)$ is mapped onto
the Thermo-Field-Double (TFD) state at infinite temperature
\be
|I\rangle \equiv\sum_k |k\rangle \otimes |k\rangle .
\label{tfd}
\ee
This is an exact eigenstate with a zero eigenvalue describing the steady ($t \to \infty$ limit) state of the vectorized system.

\subsection{SD equations and calculation of Green's functions}
In order to carry out the mentioned vectorization, following Ref.~\cite{garcia2022e}, we first employ the Choi-Jamiolkowski isomorphism to express the density matrix operator as a vector in a space where the number of degrees of freedom are doubled. In a second step, 
we set up the Keldysh integral by relating Majoranas in this doubled Fock space to a certain time contour representing backward and forward time evolution. Vectors in the doubled space, $\psi \equiv \psi^+ \otimes \psi^-$  obey the following time evolution equation, 
\be
\frac {d\psi}{dt} = \mathcal{L} \psi,
\ee
where the vectorized Lindbladian is given by
\be
\label{eq:II_liovvi}
\mathcal{L} = -iH^+ + i  (-1)^\frac{q}{2} 
H^- - i \mu \sum_i\psi^+_i\psi^-_i - \frac{1}{2}\mu 
N.
\ee
Here, $ H^+= H(\psi^+) \otimes \id$ and $ H^-= \id \otimes H(\psi^-) $
are two copies ($\pm$) of the single site SYK Hamiltonian, Eq.~(\ref{eq:singleSYK}), with the same
probability distribution but different fermions.
For notational clarity, we are not displaying  the tensor products in the remainder of the paper.
All eigenvalues of the Lindblad operator have a non-positive real part, so that  the equilibrium state corresponds to the eigenvalue
$\lambda =0$.  In terms of eigenstates of the single  SYK Hamiltonian, $H|k\rangle = E_k |k\rangle$, its eigenstate is given by the TFD state Eq. \eref{tfd}.

The path integral for the overlap of the  initial state $\psi(t_i)$ and the
final state $\psi(t_f)$ is given by
\be
\langle\psi(t_f) | \psi(t_i)\rangle =\int D\psi e^{i S}
\ee
with action given by
\be
iS= \int_{t_i}^{t_f}dt \left (-\frac 12 \psi_i^-(t) \del_t \psi_i^- (t) - \frac 12 \psi_i^+(t) \del_t \psi_i^+(t)  +\mathcal{L} \right ),
\label{act}
\ee
where we use the notation of summing over repeated indexes. 
We will be interested in sufficiently long times where the system is close to a stationary state. Given the quantum chaotic nature of the dynamics of the SYK model,
we can safely neglect any dependence on initial conditions. Despite the fact that the system is close to stationary, we still want to evaluate real time correlation functions for any time difference.  As a result of these considerations, we set  $t_i = -\infty$,
$t_f =+\infty$. The Keldysh path integral formalism involves
  combining the integrations over $\psi^+$ and $\psi^-$ into a single integration variable
  over a doubled contour in the complex plane. 
 More specifically, the integrals over $\psi^+$ and $\psi^-$
are combined
into a single contour, $\sC=\sC^+\cup \sC^-$,
where $\sC^+$ runs from $-\infty$ to $+\infty$, namely, forward in time, and $\sC^-$ runs from $+\infty$ to $-\infty$, namely, backward in time.
Physically, the need of a backward and a forward in time direction is expected because, as mentioned earlier, the density matrix is an operator so its time evolution must include both time directions. 
On the Keldysh contour $\sC$ we use the definitions $\psi^i(t^+) = \psi_i^+(t)$ and
$\psi^i(t^-) = i\psi_i^-(t)$, where $\psi_i^+$ and $\psi_i^-$ are the
independent Majoranas introduced in the action Eq.~\eref{act}. 
Since the action in the Keldysh contour is in general defined in the complex plane, it is more natural to rewrite it as,
\begin{equation}
	\begin{split}
		iS=&-\int_\sC dz \,\psi^i(z)\partial_z \psi^i(z)
		-i\int_\sC d z\, i^{q/2} \sum_{i_1<\cdots<i_q}^N J_{i_1\cdots i_q}\psi^{i_1}(z)\cdots \psi^{i_q}(z)
		\\
		&+\frac \mu 2\int_\sC d z\, d z'\, K(z,z') \psi^i(z)\psi^i(z'),
	\end{split}
\end{equation}
where $dz=dt$ on $\sC^+$, $dz=-dt$ on $\sC^-$ and
the kernel $K(z,z') $ is defined by
\be
\int_\sC d z\, d z'\, K(z,z') \psi^i(z)\psi^i(z') = \int dt\left ( \psi^i(t^+)\psi^i(t^-)- \psi^i(t^-)\psi^i(t^+) \right ).
\ee
For convenience, we did not include the irrelevant constant $-N\mu/2$.
We note that the resulting Keldysh contour (path) integral has no constraints  related, for instance, to initial conditions.
This is another simplification that is justified because we are not interested in the short time dynamics, namely, implicitly we are assuming that the system will quickly forget about initial conditions which is typical of quantum chaotic systems. We refer to Ref.~\cite{garcia2022e} for more details on the derivation of the action.

We now average over random couplings $J_{ijkl}$ and introduce two bi-local
functions $G^{ab}(z_1,z_2)$ and $\Sigma^{ab}(z_1,z_2)$
according to
\be
\delta( G^{ab}(t_1,t_2) + i \sum_i \psi^a_i(t_1)\psi^b_i(t_2))
= \int d\Sigma e^{-\frac N2\int\!\!\!\int_\sC \Sigma^{ab}(z_1,z_2)( G^{ab}(z_1,z_2) + i \sum_i \psi^a_i(z_1)\psi^b_i(z_2))}
\ee
with labels $a$ and $b$
$\in\{+, -\}$. After integrating the fermions, the path integral $\int DGD\Sigma e^{iS}$ is a functional integral with respect to $G$ and
$\Sigma$ which can be evaluated by a saddle-point approximation. The Schwinger-Dyson equations are given by (see Appendix \ref{app:kernel}), 
\be
(i\delta(z-z')\del_z -\Sigma(z,z'))G(z,z') &=& \mathrm{1}_\sC,\\
\Sigma(z,z') +i^q J^2 [G(z,z')]^{q-1} -\frac i2 \mu[ K(z,z') -K(z',z)] &=& 0,
\ee
where we have used that $G(z,z')=-G(z',z)$. 
In terms of the $\pm$ components and the time variables, these equations can be written down as
\be
\left ( \begin{array}{cc}
 (\i \del_t-\Sigma^{++})G^{++} +\Sigma^{+-} G^{-+} &  (\i \del_t-\Sigma^{++})G^{+-} +
\Sigma^{+-} G^{--} \\
-(\i \del_t+\Sigma^{--})G^{-+} +\Sigma^{- +} G^{++}
&  -(\i \del_t+\Sigma^{--})G^{--} + \Sigma^{-+} G^{+-} \end{array}\right )
 =\left(\begin{aligned}
\delta(t,t')  & \quad 0 \\  
0 \quad & \delta(t,t') 
\label{eq:SD1}
\end{aligned}\right),\ee 
and
\begin{equation}\begin{aligned}
    & \Sigma^{++}(t,t')+J^2 i^q G^{++}(t,t')^{q-1}=0,
    & \Sigma^{--}(t,t')+J^2 i^q G^{- -}(t,t')^{q-1}=0,\hspace{5em} \\ 
    & \Sigma^{+-}(t,t') +J^2 i^q G^{+-}(t,t')^{q-1} -i\mu\delta(t,t')=0,
    & \Sigma^{-+}(t,t') +J^2 i^q G^{-+}(t,t')^{q-1} +i\mu\delta(t,t')=0. 
\label{eq:SD2}
\end{aligned}\end{equation}
It is often convenient to express the  solutions of the SD equations above
 in terms of  the real time Green's functions
\be
G^<(t,t') &=& \langle G^{+-}(t,t') \rangle\nn, \\ 
G^>(t,t') &=& \langle G^{-+}(t,t') \rangle\nn, \\ 
G^T(t,t')&=&\langle G^{++}(t,t') \rangle,\\
G^{\oT}(t,t')&=&\langle G^{--}(t,t') \rangle.
\ee
Similar combinations apply to the $\Sigma$ variables.

Because the Keldysh contour is time ordered, we have that
\be
G^T(t,t')&=& \theta(t-t') G^>(t,t') + \theta(t'-t) G^<(t,t'),\\
G^\oT(t,t')&=& \theta(t-t') G^<(t,t') + \theta(t'-t) G^>(t,t').
\ee
This gives the identity
\be
G^T(t,t') +G^\oT(t,t')=G^>(t,t')+G^<(t,t').
\label{cond1}
\ee
The action is invariant under $\psi(t^+) \to \psi(t^-)$,  $\psi(t^-) \to \psi(t^+)$ and $i \to -i$.
Therefore,
\be
G^>(t,t') = -G^<(t,t'),
\label{cond2}
\ee
or, in Fourier space, $G^>(\omega) = - G^<(\omega)$. Taking into account the usual relation for finite temperature Green's functions, $G^>(\omega) = - \exp(\beta\omega) G^<(\omega)$,
we conclude that the Green's functions are at infinite temperature, which is consistent
with the infinite temperature equilibrium state of the Lindblad evolution. In appendix
\ref{app:sd}, we will  show that the conditions \eref{cond1} and \eref{cond2}
also follow from the SD equations.

The Fourier transformed SD equations are given by
\be
(\omega- \Sigma^T(\omega))G^T(\omega) + \Sigma^<(\omega) G^> (\omega) &=&1,\\
(\omega- \Sigma^T(\omega))G^<(\omega) + \Sigma^<(\omega) G^{\overline{T}} (\omega) &=&0,\\
(\omega+\Sigma^{\overline{T}}(\omega))G^>(\omega) -\Sigma^>(\omega) G^{{T}} (\omega) &=&0,\\
-(\omega+ \Sigma^{\overline{T}}(\omega))G^{\overline{T}}(\omega)
+\Sigma^>(\omega) G^{{<}} (\omega) &=&1.
\label{cont}
\ee
Generally, the solution of the SD equations requires an $i\epsilon$ prescription,
$\omega \to \omega +i\epsilon$, but for given boundary conditions, 
a nonzero value of $\mu$ already selects solutions
with the correct causal structure. 

In order to obtain the Lyapunov exponent, our first task is the evaluation of the retarded Green function $G^R(t) = \theta(t)(G^>(t)-G^<(t))$. By using that the solutions of the
SD equations satisfy the conditions \eref{cond1} and \eref{cond2},
the third and fourth SD equations coincide with the first two.
The difference and the sum of the
    first  two SD
    equations results in
\be
(\omega - \Sigma^R(\omega) )G^R(\omega) =1,\qquad
(\omega - \Sigma^A(\omega) )G^A(\omega) =1.
\ee
where we have defined the retarded and advanced Green's function,
    \be
G^R(\omega) &=& G^T(\omega) - G^<(\omega) ,\\
G^A(\omega) &=& G^T(\omega) + G^<(\omega),
\ee
with analogous expressions for $\Sigma^{R,A}$.
To solve the SD equations, we will employ the reality conditions of the
Green's functions which we will derive next.
Following arguments in Ref.~\cite{garcia2022b} for a closely related model,
they can also be obtained from the
symmetry properties of the action.

The Euclidean time ordered Green's function
(denoted by a superscript  $E$) in the limit of interest is given by \cite{nosaka2021}
\be
G^E(u,u') =  \left \langle \frac 1N \Tr \sum_i \mathcal{T}  \psi^i(u) \psi^i(u') \right \rangle, \label{euclideang}
\ee
with $\mathcal{T}$ the time ordering operator, and satisfies the analyticity requirement \cite{nosaka2021}
\be
{G^E}^*(u,u') = - G^E(-u^*,-{u'}^*).
\label{ana1}
\ee
If the Hamiltonian is non-Hermitian, this identity  is only valid
if $H^\dagger$ and $H$ have
the same weight in the ensemble of random SYK models.
Therefore,
\be
   {G^>}^*(\omega) &=& - {G^>}(\omega) =  {G^<}(\omega),\nn\\
  { G^T}^*(\omega) &=& G^T(\omega) = -G^{\oT}(\omega),
\ee
and
\be
G^>(\omega) = - G^<(\omega)
=\frac 12 ({G^R(\omega) - G^R(\omega)^*}).
\ee
We remind the reader that throughout this paper $\beta = 0$.
In terms of the real time Green's functions, the SD equations can be simplified to
\be
   \nonumber G^<(\omega) &=& - i{\Im G^R(\omega)},\\ \nonumber
    \Sigma^R(\omega) +i\mu&=& -2i J^2 \int_0^{+\infty}dt e^{i(\omega+i\epsilon)t}
    \mathrm{Im}[G^<(t)^3], \\
    G^R(\omega) &=&\frac{1}{(\omega + i\epsilon) -\Sigma^R(\omega)},
\label{real-time}
\ee
where we have used that $\Sigma^R(t) =\theta(t) (\Sigma^>(t)-\Sigma^<(t) )$,
Eq. \eref{eq:SD2},  $G^>(t) = (G^<(t))^*$, 
and $\int dt \delta(t) \theta(t) = 1/2$.
The real time equations above are solved by standard iterative methods. 
Strictly speaking, these expressions are only applicable to systems close to thermodynamical equilibrium. As mentioned earlier, the steady state resulting from the Lindbladian dynamics is \cite{garcia2022e} a TFD at infinite temperature $(\beta = 0)$.  At sufficiently short times, the dynamics will depend on the initial state and therefore cannot be described by the SD equations which were derived from the Keldysh path integral without introducing any initial state constraint or specifying a specific quench protocol. This independence on initial conditions is anticipated, even for relatively short time scales, because classically chaotic systems lose memory of the initial conditions rather quickly, so this loss occurs much earlier than the exponential growth of the OTOC around the Ehrenfest time due to the proliferation of quantum effects. However, for some choices of initial conditions, the loss of memory may take longer to occur which may lead to non-universal deviations from the OTOC results reported below.  
In any case, we expect this approximation to be valid in most cases, especially for initial states given by a TFD state at sufficiently high temperature.  

In the next section, we proceed with the computation of the OTOC around the Ehrenfest time for an SYK coupled to a Markovian bath.  

\section{Calculation of the OTOC}
After solving the SD equations and finding an explicit expression for $G^R$, we show in appendix \ref{app:kernel} that the Lyapunov exponent $\lambda_L$ is computed by solving
\begin{equation}\begin{aligned}
\mathcal{F}(t_1,t_2)=\int dt_3 dt_4 K_R(t_1,t_2;t_3,t_4)\mathcal{F}(t_3,t_4), 
\label{eq:ladder}
\end{aligned}\end{equation}
using the Ansatz
\be
 \mathcal{F}(t_1,t_2) =e^{\lambda_L(t_1+t_2)/2}f(t_1-t_2) .
\label{ansatz}
 \ee
 The Bethe-Salpeter equation \eref{eq:ladder}
is derived from the path integral formalism by evaluating the leading $1/N$ correction and expressing it in terms of Green's functions resulting from the solution of the SD equations.
It turns out that the expression for the kernel $K_R(t_1,t_2;t_3,t_4)$ has the same structure
as the one for the single site SYK model \cite{maldacena2016}, 
\begin{equation}\begin{aligned}
K_R(t_1,t_2;t_3,t_4) = -(q-1)J^2 G^R(t_1-t_3)G^R(t_2-t_4)  G_{W}^{q-2}(t_3-t_4), \\
\end{aligned}\end{equation}
where $G_{W}$ is the Wightman function, derived in appendix \ref{app:kernel},  which can be expressed as the analytic continuation of $G^E$, the Euclidean Green's function (\ref{euclideang}), $G_{W}(t)= G^E(it+\beta/2)$ with $\beta = 0$ in this case. 

It is in principle surprising that the structure of this kernel function is the same as the one in the original SYK model \cite{maldacena2016} even though the SYK model is now coupled to a Markovian bath. The reason is that the specific choice of jump operators we are considering, linear in the Majorana fermions, leads to a coupling term in the path integral which is linear in $G^>$ and $G^<$. As a result, the leading perturbations to the large $N$ limit that are necessary for the calculation of the OTOC only contribute linearly.
Since the derivation of the kernel function relies on second-order perturbations to the Green's functions, the form of the kernel function does not depend on the coupling $\mu$, though the OTOC does depend on it through the retarded Green's function and the Wightman functions. More details can be found in appendix \ref{app:kernel}. We stress that this independence of the bath is a particularity of our choice of jump operators. In general, jump operators involving two or more Majorana fermions will also modify the structure of the kernel equation.  

Using the Ansatz \eref{ansatz},
 the kernel equation \eref{eq:ladder} can be transformed into
 \be
 f(t_1-t_2)&=&- (q-1)J^2\int_{-\infty}^{t_1}\int_{-\infty}^{t_2} dt_3dt_4 e^{\frac{\lambda_L}2(t_3+t_4-t_1-t_2)}\nn \\ &&\times
G^R(t_1-t_3)G^R(t_2-t_4) G_W^{q-2}(t_3-t_4) f(t_3-t_4). \nn \\
\ee
At  time scales where the system is close to the equilibrium state, the retarded
Green's function is expected to simplify to
\be
G_R(t) \approx  -i\theta(t) e^{-\Gamma t},
  \ee
  and the ladder  equation reduces to a differential equation
  \be
  ( \del_{t_1} + \frac{\lambda_L}2+\Gamma) ( \del_{t_2} + \frac{\lambda_L}2+\Gamma)f(t_1-t_2)
  = (q-1)J^2 G_W^{q-2}(t_1-t_2) f(t_1-t_2).
  \ee
  Using translational invariance, this becomes the Schr\"odinger-like equation
  \be
 \left[ -\del_t^2 +\left (\frac{\lambda_L}2+\Gamma\right )^2- (q-1)J^2 G_W^{q-2}(t)\right ] f(t) =0
 \ee
 of a particle in the potential $-(q-1) J^2 G_W^{q-2}(t)$.  Note that $G_W(t)$ is an even function of
 $t$. 
 In the next section, we will analytically  solve this equation in the large $q$ limit. In the limit $\beta \to 0$ of interest, we would have to solve the real time equations \eref{real-time}. In that
case, the Fourier transformed Wightman function is obtained from the Euclidean
Green's function \cite{nosaka2021}
\be
G^E(u) = -
\int \frac {d\omega}{2\pi} e^{-\omega u} \mathrm{Im}[G^R(\omega)]
\ee
resulting in the Wightman function $G^W(t) =G^E(it)$, and
\be
 G_{W}(\omega) ={\rm Im}[ G^R(\omega)].
\label{gw1}
 \ee
 The function $f(\omega)$ obeys the integral equation 
 \begin{equation}\begin{aligned}
f(\omega)= (q-1)J^2\left | G^R(\omega+i\frac{\lambda_L}{2})\right|^2 \int \frac{d\omega'}{2\pi}g_{W}(\omega-\omega')f(\omega'),
\label{eq:kernel}
 \end{aligned}\end{equation}
 where $g_W(\omega) $ is the Fourier transform of $G_W^{q-2}(t)$.
   Numerically, 
   after discretizing $\omega$, this equation becomes a finite dimensional eigenvalue equation. Its eigenvalues depend on $\lambda_L$, and $\lambda_L$ is determined by the condition that the integral equation has an eigenvalue 1. Numerically, this can be done
   efficiently by bi-section in $\lambda_L$.

\begin{figure}[htbp]
	\centering
	\includegraphics[scale=.9]{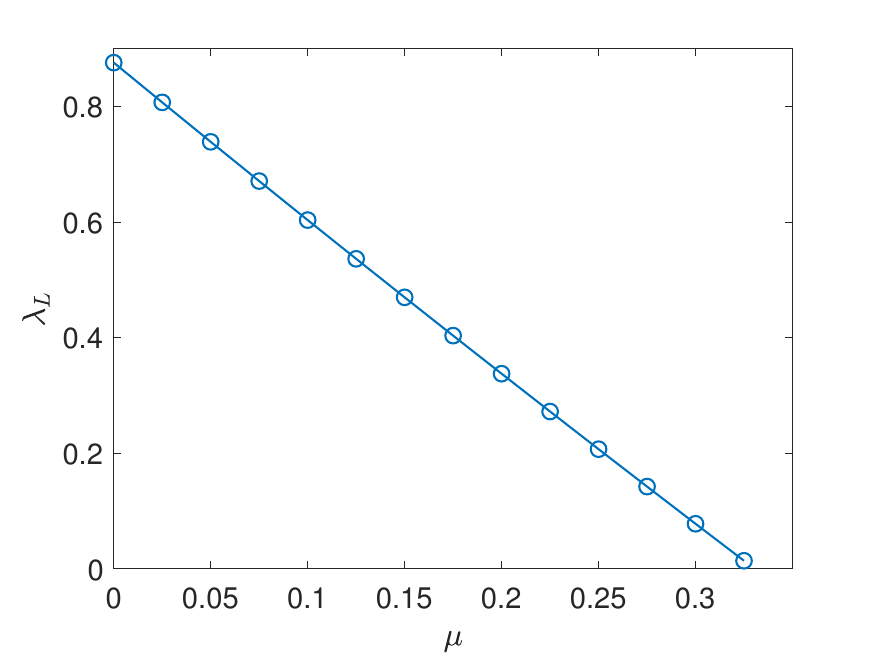}
	\caption{The Lyapunov exponent $\lambda_L$ as a function of the coupling to the bath $\mu$ from the numerical solution of the ladder equation (\ref{eq:ladder}) for $q = 4$. We observe that $\lambda_L$ decreases with $\mu$ and eventually becomes negative for $\mu > 0.33$. }\label{fig:lyp_lambda_beta_0}
\end{figure}

\subsection{Numerical OTOC and Lyapunov exponent}\label{sec:otoc}
The SD equations are solved numerically by iterating the real-time equations \eref{real-time}.
After a Fourier transform, this gives the retarded Green's function $G^R(\omega)$. The Wightman function
$G_W(\omega) $
is obtained from $G^R(\omega)$ according to \eref{gw1}. This allows us to solve the kernel equation, also called the Bethe-Salpeter 
equation \eref{eq:kernel}, by discretizing the frequencies and numerically solving the ensuing eigenvalue
equation.

It was shown in Ref.~\cite{garcia2022e} that for sufficiently long times and $\mu > 0$,
the steady state resulting from the Liouvillian dynamics is a TFD state at infinite temperature ($\beta = 0$).
Following the numerical procedure explained above,  we show in Fig. \ref{fig:lyp_lambda_beta_0} the Lyapunov exponent $\lambda_L$ as a function of the coupling to the bath for $q=4$. Interestingly, it shows a monotonous, almost linear, decrease of the Lyapunov exponent with the coupling to the bath $\mu$ until it  becomes negative for $\mu > \mu_c \approx 0.33$.
A negative value of  $\lambda_L$ indicates  that the exponential growth of the OTOC, which physically describes the increase in quantum uncertainty with time,
has been fully suppressed by the coupling to the bath. It is tempting to identify this point as the
transition between quantum chaotic dynamics, mostly driven by the single-site SYK model, and non-chaotic dynamics mostly driven by the environment. More work
needs to be done to fully characterize the nature of this transition.

\subsection{Analytic calculation of $\lambda_L$ in the large $q$ limit}\label{sec:otoclargeq}

We now proceed to the analytic calculation of the Lyapunov exponent
for $\beta \to 0$ in the large $q$ limit
at fixed
\begin{align}
	\mathcal{J}^2 = \frac{q J^2}{2^{q-1}}\quad{\rm and} \quad  \hat \mu = \mu q.
\label{cal-J}
\end{align}
In this limit, the self-energy becomes $O(1/q)$ which makes it possible to solve the SD equations analytically. Following the derivation for the Maldacena-Qi model \cite{maldacena2018},
the  real time Green's functions can be expanded as, 
\begin{equation}\begin{aligned}
    & G^T(t)=-\frac{i}{2}\mathrm{\sgn(t)} (1+\frac{1}{q}g^{T}(t)+\cdots), \\
    & G^{<}(t)=\frac{i}{2} (1+\frac{1}{q}g^{<}(t)+\cdots),
    \label{gq}
\end{aligned}\end{equation}
and, for $\beta = 0$,  $G^{>} = -G^{<}$.
Therefore, 
\begin{equation}\begin{aligned}
G^R(t) = \theta(t)(G^{>}(t)-G^{<}(t))=-i\theta(t) (1+\frac{1}{q}g^{<}(t)+\cdots),
\label{gr1q}
\end{aligned}\end{equation}
and the self-energies are given by
\be
\Sigma^T(t)&=& -i^q J^2 {G^T(t)}^{q-1} =\frac{i^{2q-1} J^2}{2^{q-1}}\sgn^{q-1}(t)e^{g^T(t)},\nn\\
    \Sigma^<(t)&=& -i^q J^2 {G^{<}(t)}^{q-1}+i\mu =\frac{i^{2q-1} J^2}{2^{q-1}}  e^{g^<(t)} +i\mu.
      \label{sigq}
      \ee
      Substituting the large $q$ expressions for $G$ and $\Sigma$ into the Schwinger-Dyson equations
      and collecting the $1/q$ contributions (the leading order cancels) results in the differential equations
      \be
     && \del_t^2 (\sgn(t) g^T(t)) + 2\mathcal{J}^2 \sgn(t) e^{g^T(t)}=0,\nn\\
      &&\del_t^2g^<(t) + 2\mathcal{J}^2 e^{g^<(t)}=0.
        \ee
        The solutions of these equations are given by
        \be
        e^{g^T(t)}=\frac{\alpha^2}{\mathcal{J}^2 \cosh^2(\alpha|t|+\gamma)},\qquad
          e^{g^<(t)}=\frac{\tilde{\alpha}^2}{\mathcal{J}^2 \cosh^2(\widetilde{\alpha}|t|+\widetilde{\gamma})},
\label{eq:solutions_q}
\ee
        where $ \alpha,\;\widetilde \alpha, \; \gamma$ and $\widetilde \gamma$ are integration constants
        determined by the conditions 
\begin{equation}\begin{aligned}
    \frac{\alpha^2}{\mathcal{J}^2 \cosh^2 \gamma}=1,\quad
    2\tilde{\alpha}\tanh\tilde{\gamma}=\hat{\mu},\quad \tilde{\alpha}
    =\alpha, \quad \tilde{\gamma}=\gamma,
\label{eq:constants_0}
\end{aligned}\end{equation}
resulting from the imposition of the following boundary conditions: $G^T(0^+)={i}/{2}$, so $\exp({g^T(0)})=1$, $\partial_t g^{<}(0)=-\hat{\mu}$, which can be derived by recalling the Schwinger-Dyson equations and applying the ${1}/{q}$ expansion.
The last two conditions in \eref{eq:constants_0} 
come from the relation $G^T(t)=-\mathrm{\sgn}(t) G^<(t)$ in the $t \to \infty$ limit \cite{maldacena2018}.

For large $t$, the retarded Green function $G^R(t)$ is expected to decay exponentially,
$G^R(t) \propto e^{- \Gamma t}$. This can be shown explicitly in the limits of large 
$q$ and $t\gg 1$ \cite{kawabata2022,maldacena2018,khramtsov2021,garcia2022e}.
Using that the Green's function vanishes for $t\to \infty$ as 
\be
G^<(t) = \frac i2   e^{-2\alpha t/q},
\ee
where  the exponent is obtained by matching to the $t \ll q$ solution. Although this result
is only valid for $t \gg 1$, it also turns out to be a good approximation for $t <1$
\cite{Tarnopolsky:2018env}. To enable a fully analytical calculation of the Lyapunov exponent,
we use this approximation so that the retarded Green's function is given by
\be
G^R(t) = -i \theta(t) e^{-2\alpha t/q}.
  \ee
  The same result can be obtained by a na\"{\i}ve exponentiation of  
  the $1/q$ correction in \eref{gr1q}.
  The accuracy of this approximation has been confirmed by comparing with results from the numerical solution of the SD equations at a large but finite $q$. We note that this approximation is incompatible with a strict large $q$ expansion
  but, at finite $q$, as in the Hermitian case \cite{maldacena2016}, we shall see it gives a much better approximation.
  For example, in the weak-coupling/high-temperature limit of the single site SYK model, the $1/q$-correction to the Lyapunov exponent is given by
  $2 \mathcal{J}(1 -2/q)$ 
  so approximately for $q = 4$ 
  the Lyapunov exponent is suppressed by a factor two in rough agreement with numerical results from
  solving the SD equations (see Fig. 11 of \cite{maldacena2016}).

  The $(q-2)$th power of the Wightman function (see Eq. \eref{gw1} for an expression of the
  Wightman function in terms of the retarded Green's function)
for $\beta \to 0$ is  given by, 
\begin{align}
  & (q-1) J^2 G_{W}^{q-2}(t) =  (q-1) J^2(-i)^{q-2}{G^<}^{q-2}(t)
= \frac {2\alpha^2}
{ \cosh^2(\alpha|t|+\gamma)}.
\label{gw}
\end{align}
Therefore, in the kernel equation \eref{eq:ladder}, the main contributions to the integral come from
$t_3-t_4 \sim O(1) $ and both  $ t_1- t_3$ and $t_2 -t_4$  $\sim O(q)$.  

From the discussion in the previous section, we know that
for an exponentially decreasing retarded Green's function the
 kernel equation reduces to the differential equation
\begin{align}
	& \left [-\partial_t^2 +\left (\Gamma+\frac{\lambda_L}2 \right )^2  - \frac{2\alpha^2}{ \cosh^2(\alpha|t|+\gamma)} \right]f(t) =0. 
\label{eq:ladder_derivative}
\end{align}
This is a Schr\"odinger-like equation for which we are seeking bound state solutions.
 Since
the potential has a discontinuous derivative at $t=0$, we have to separate
solutions for $t<0$ (equivalent to $\gamma \to -\gamma$) and $t>0$.  In order to satisfy the differential equation,
these two solutions must be continuous and with a
continuous derivative at $t = 0$.
The general solution of this differential equation
is given by, 
\be
f(t)=c_1 e^{-\sigma (t+\gamma/\alpha)}
(\sigma + \alpha \tanh(\alpha t+\gamma))
+c_2  e^{\sigma (t-\gamma/\alpha)} (-\sigma
+\alpha \tanh(\alpha t -\gamma)).
    \ee
    Using the boundary conditions that $f(|t| \to \infty) =0$, we have that $c_2=0$
    for $t>0$ and $c_1=0$ for $t<0$ when also $ \gamma \to -\gamma$. Continuity
    of the solution at $t =0 $ requires that $c_2 = -c_1$. We thus find
    the solution
    \be
    f(t)&=& c_1\theta(t) e^{-\sigma (t+\gamma/\alpha)} (\sigma + \alpha \tanh(\alpha (t+\gamma/\alpha))
      -c_1 \theta(-t)  e^{\sigma (t-\gamma/\alpha)} (-\sigma +\alpha \tanh(\alpha (t-\gamma/\alpha))\nn\\
      &=& c_1 e^{-\sigma (|t|)+\gamma/\alpha) }
      (\sigma + \alpha \tanh(\alpha |t|+\gamma)).
        \label{solution}
        \ee
        The second derivative in Eq.\eref{eq:ladder_derivative}
        gives rise to a term $f'(t) \delta(t)$, which has to vanish for $f(t)$ to be
        a solution of the differential equation \eref{eq:ladder_derivative},  i.e.
        $f'(0)=0$.        Therefore,
        \be
        \frac \sigma \alpha \left (-\frac \sigma\alpha -\tanh \gamma)\right )
        + \frac{1}{\cosh^2\gamma} =0.
        \ee
        Solving for ${\sigma} /{\alpha}$ leads to,
        \be
        \frac {\sigma}{\alpha} = -\frac 12 \tanh \gamma +\frac
              { \sqrt{4+ \sinh^2\gamma}}{2 \cosh \gamma}.
              \label{eq:nu_exp_gamma}
              \ee
              From Eq.~(\ref{eq:constants_0}) we obtain
              \be
              \sinh \gamma &=& \frac{\hat \mu}{2 \mathcal{J}}, \qquad
              \alpha = \mathcal{J} \cosh \gamma.
              \ee
              Combining these equations with the expression for
              $\sigma/\alpha$ (\ref{eq:nu_exp_gamma}), we can solve for the Lyapunov exponent $\lambda_L$ and the
              decay rate $\Gamma$ as a function of $\mu$, 
              \be
              \lambda_L =2\sigma -\frac{4\alpha}q, \qquad
              \Gamma \equiv\frac{ 2\alpha}q. \label{eq:lyalargeq}
              \ee
        \begin{figure}[t!]
          \includegraphics[width=10cm]{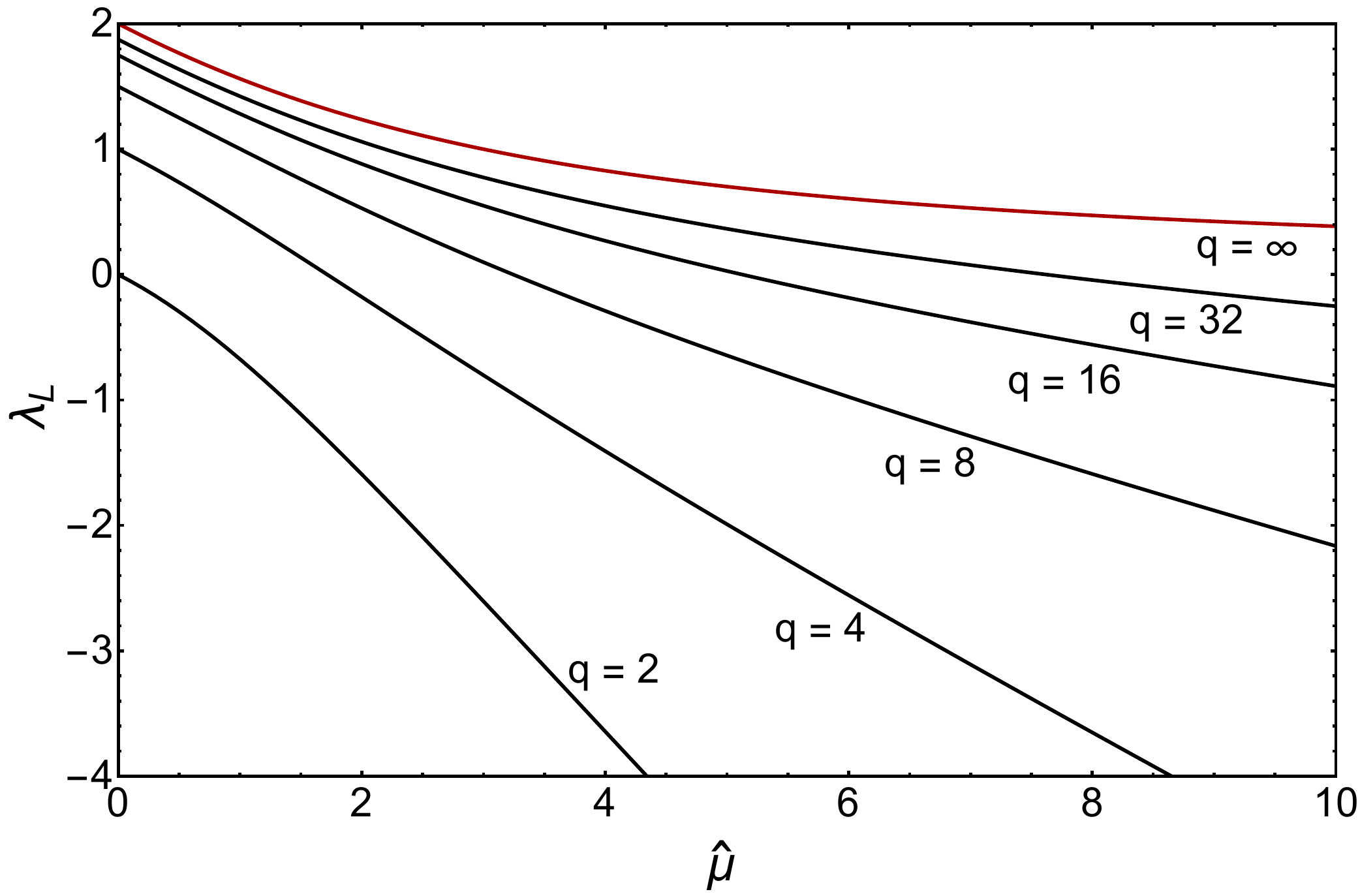}
          \caption{Analytic Lyapunov exponent $\lambda_L$ as a function $\widehat \mu =\mu q$ for
            various finite values of $q$ (black curves) and $q\to \infty$ (red curve). It becomes negative for sufficiently large ${\hat \mu}$.
            \label{fig:lyaa}}
.          \end{figure}
              In Fig. \ref{fig:lyaa}, we show the result of the Lyapunov exponent versus $\widehat \mu$ for various values of $q$. Beyond a critical value of $\mu$, the Lyapunov exponent becomes negative which indicates the strong suppression of quantum effects consistent with a transition from quantum chaotic to quasi-classical dynamics. We note that although these are results from a large-$q$ analysis, the Lyapunov exponent for $q = 2$ still vanishes as was expected for a non quantum chaotic system.

\begin{figure}[htbp]
	\centering
	\subfigure[]{\includegraphics[scale=.50]{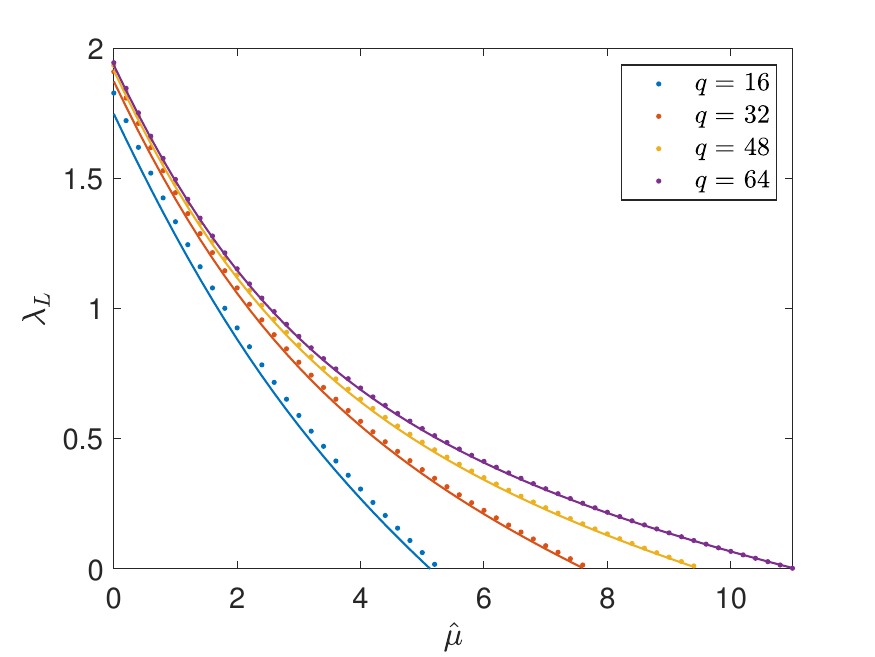}}
	\subfigure[]{\includegraphics[scale=.50]{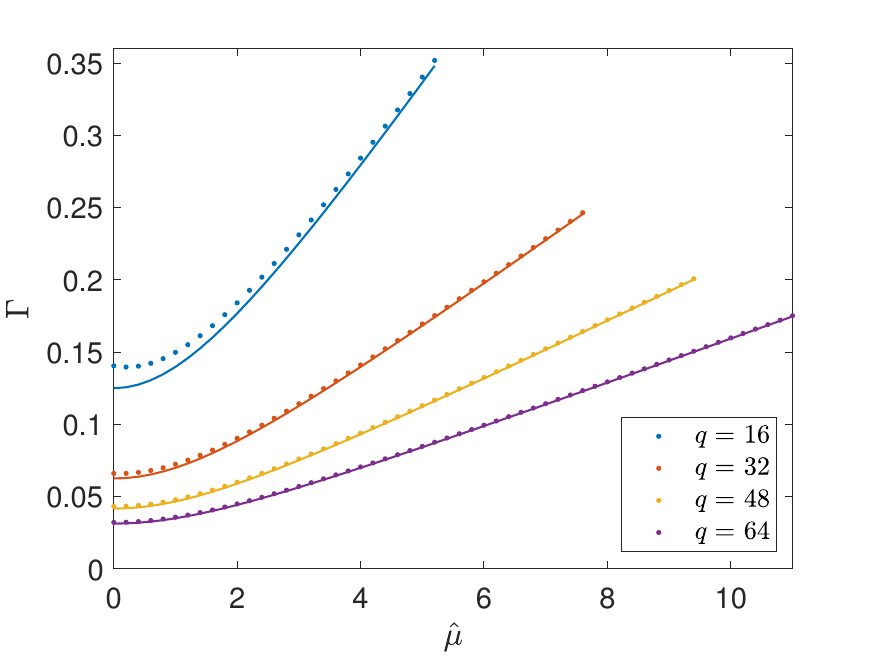}}
	\caption{Lyapunov exponent $\lambda_L$ (left) and decay rate $\Gamma$ (right) for $\mathcal{J} = 1$, $q =~16,~32,~48,~64$, $\beta = 0$ for various $\hat{\mu}$. The dots
          correspond to the numerical solution of the kernel equation (\ref{eq:kernel}), while the solid curves  stand for the analytic large $q$ prediction Eq.~(\ref{eq:lyalargeq}). We observe that the agreement becomes better as $q$ increases. They are already almost indistinguishable for $q = 64$.  }\label{fig:lyp_gm_ld_q_16_64_J_cal_1_T_infty_anlyt}
\end{figure}

              The resulting analytical expressions for $\Gamma$ and $\lambda_L$, together with
              the numerical results for $q = 16,\; 24, \;32$ and $64$, from solving the SD equations numerically, 
              are shown in Fig.~\ref{fig:lyp_gm_ld_q_16_64_J_cal_1_T_infty_anlyt}.
The value of the coupling constant is equal to  $\mathcal{J}= \sqrt{q}{J}/{2^{{(q-1)}/{2}}} = 1$. 
The agreement with the analytical result improves for increasing values of $q$. Also
when
$\lambda_L <0$, the SD equations and the Bethe-Salpeter equations can be solved in the same way without any problem. The  results show a  similar  agreement with the analytical results, but we do not
show this part of the curves. 

We have also checked that the solution, Eq. \eref{solution}, satisfies the
integral equation \eref{eq:ladder_derivative} by an explicit evaluation
of the integral (see appendix \ref{app:solkerb0}). In that case, the quantization
condition \eref{eq:nu_exp_gamma} results from imposing that the solution
Eq. \eref{solution} is an eigenfunction of the integral kernel equation (\ref{eq:ladder_derivative}).

\section{Discussion and Conclusions}\label{sec:conclusions}
The results of this paper  shed light on  the question of precisely defining quantum chaos in dissipative systems. 
In principle, it is tempting to extend the BGS conjecture to systems with a complex spectrum,
namely to state that dissipative quantum chaotic dynamics is characterized by the agreement
of the spectral correlations of the system, in our case an SYK model coupled to a bath, with the predictions
of an ensemble of non-Hermitian random matrices with the corresponding symmetries. Indeed,
at least in certain cases, it has been found \cite{garcia2022d,li2021a} level statistics of
non-Hermitian many-body quantum systems are well described by non-Hermitian random matrix theory.
However, dissipation tends to suppress quantum effects, so it is rather unclear whether the
extension of  the BGS conjecture to non-Hermitian systems is in all cases meaningful, even if
spectral correlations are well described by (non-Hermitian) random matrix theory. The findings
of this paper show that dissipation suppresses the Lyapunov exponent and eventually leads to
a negative Lyapunov exponent for a sufficiently strong dissipation strength.  In that case,
quantum fluctuations decrease with time, so the initial wave packet stays close to the
classical trajectory. This suggests that a necessary condition for the existence of dissipative $quantum$ chaos
is a positive Lyapunov exponent. However, a warning is in order, we have studied a very specific
Markovian bath leading to a coupling term in the Liouvillian which is both integrable and Hermitian.
It would be necessary to confirm whether the observed transition occurs for more general jump
operators with a larger number of Majorana fermions and random couplings \cite{sa2022,kawabata2022} leading to coupling terms in the Liouvillian which are both non-Hermitian and potentially quantum chaotic.  

Another issue, also related to the definition of quantum chaos in Hermitian systems, is 
the bound \cite{maldacena2015} on the Lyapunov exponent $\lambda_L \leq 2\pi T/\hbar$ for systems
at finite temperature $T$. The saturation of the bound is a signature of many-body quantum chaotic
systems with a gravity dual. It would be interesting to explore the extension of the bound
to dissipative quantum chaos.  
Unfortunately, our setting leads to a steady state at infinite temperature whose Lyapunov exponent is not restricted by the bound. Different jump operators will
generally lead to different steady states so fine tuning may be necessary to equilibrate
to a Gibbs state which complicates the study of finite temperature effects.

Finally, we comment on the universality of our results. As was mentioned in the introduction,
a similar vanishing of the Lyapunov exponent has been reported \cite{chen2017} in a two-site SYK model
where one of the SYK models acts as a thermal bath provided that the coupling to the system,
the other SYK, is strong enough. We do not believe that this result depends on the details
of the SYK model. Quite the contrary, to some extent, the SYK model is the most quantum
chaotic system due to the random couplings and the infinite range of its interactions,
so we would expect that interacting models with a sparser interaction will be less robust to the
suppression of quantum effects due to the contact with a bath and the subsequent
suppression of the exponential growth of the OTOC when the coupling is sufficiently strong.
However, as mentioned earlier, additional results within the Lindbladian approach, using more general jump operators, are certainly desirable to confirm this expectation. 
 
In conclusion, we have studied the OTOC for the vectorized formulation of
a single-site SYK model coupled to a Markovian bath described
by the Lindblad formalism. We have found that the growth is still exponential around the Ehrenfest time,
but the Lyapunov exponent decreases with the coupling strength to the bath and eventually becomes negative leading to a transition whose details remain to be studied. Therefore, a natural definition of quantum chaos in dissipative systems is the existence of a positive Lyapunov exponent independent of whether level statistics agrees with the corresponding non-Hermitian random matrix prediction whose precise relation to dissipative quantum chaos dynamics remains to be demonstrated. We note that a limitation of this characterization is that it requires a well defined semiclassical limit, in the SYK model it is the existence of a $1/N$ expansion. In certain Hermitian systems, like certain spin-chains \cite{lin2018,craps2020,jalabert2018,garciamata2023} whose level statistics is well described by random matrix theory \cite{luitz2015}, no exponential growth of the OTOC is observed because of this reason. 
 
As mentioned earlier, a natural extension of this work is to study the dynamics for more general jump
operators in order to clarify the details of the transition.
It would also be interesting to compute the Lyapunov exponent in dissipative spin-chains
and other similar settings in order to test the universality of the suppression of chaos in dissipative
systems.
Moreover, it would be worthwhile to extend our results to finite temperature and to work out the gravity dual interpretation,
especially the mentioned transition. We note that our SYK setting has striking similarities with a wormhole configuration in a global near de Sitter background in two
dimensions \cite{turiaci2019,cotler2020,garcia2022e,shiu2020}. 
\acknowledgments{
  We thank Lucas Sa for interesting discussions. 
  A. M. G. G. acknowledges illuminating correspondence with Victor Godet.
  A. M. G. G. and J. Z. acknowledge support from the
  National Natural Science Foundation of China (NSFC):
  Individual Grant No. 12374138, Research Fund for
  International Senior Scientists No. 12350710180, and National
  Key R$\&$D Program of China (Project ID:
  2019YFA0308603). A. M. G. G. acknowledges support
  from a Shanghai talent program.
   J. J. M. V. is
  supported in part by U.S. DOE Grant  No. DE-FAG-88FR40388. 
}

\appendix
\section{Conventions}
In this appendix, we summarize some of the conventions we have been using in this paper.

The Fourier transform and its inverse are defined by
\be
f(t)& =& \int_{-\infty}^\infty \frac{d\omega}{2\pi} e^{-i\omega t} f(\omega),\\
f(\omega) &=& \int_{-\infty}^\infty dt e^{i\omega t}f(t).
\ee
For translational invariant functions we write
\be
f(t_1, t_2) = f(t_1-t_2) = f(t)
\ee.

\section{Solution of the Schwinger-Dyson Equations}
\label{app:sd}
In this appendix, we show that the solutions of the SD equations satisfy the
properites
\be
G^>(\omega) &=& - G^<(\omega),\nn\\
G^{T}(\omega) &=& - G^\oT(\omega).
  \ee

 To solve the SD equations we introduce
\be
\widetilde \Sigma^{+-}(t,t') &=& \Sigma^{+-}(t,t') -i\mu \delta(t,t'),\nn\\
\widetilde \Sigma^{-+}(t,t') &=& \Sigma^{-+}(t,t') +i\mu \delta(t,t').
  \ee
  Then the  $\widetilde \Sigma^{\alpha \beta}$, with $\alpha, \beta \in \{+,-\}$, have the same transformation properties
  under $t \to -t$ as the corresponding Green's functions. For the Fourier transformed
  Green's functions and self-energies, this results in
  \be
  \widetilde \Sigma^>(-\omega) =  - \widetilde \Sigma^<(\omega).
  \ee
  We thus find the SD equations
\be
(\omega- \Sigma^T(\omega))G^T(\omega) + (\wSigma^<(\omega)+i\mu) G^> (\omega) &=&1,\\
(\omega- \Sigma^T(\omega))G^<(\omega) + (\wSigma^<(\omega) +i\mu)G^{\overline{T}} (\omega) &=&0,\\
(\omega+\Sigma^{\overline{T}}(\omega))G^>(\omega) -(\wSigma^>(\omega)-i\mu) G^{{T}} (\omega) &=&0,\\
-(\omega+ \Sigma^{\overline{T}}(\omega))G^{\overline{T}}(\omega)
+(\wSigma^>(\omega)-i\mu) G^{{<}} (\omega) &=&1.
\label{cont1}
\ee
Adding the second and third equation  and subtracting the fourth equation from
the first one gives
\be
\label{a1}
(\omega- \Sigma^T(\omega))G^<(\omega)
+ \wSigma^<(\omega) G^{\overline{T}} (\omega) +
(\omega+\Sigma^{\overline{T}}(\omega))G^>(\omega) -\wSigma^>(\omega) G^{{T}} (\omega) &=& -i\mu (G^{\overline{T}} (\omega) + G^{{\oT}} (\omega) ),\nn \\
\label{a2}\\
(\omega- \Sigma^T(\omega))G^T(\omega) + \wSigma^<(\omega) G^> (\omega)
+(\omega+ \Sigma^{\overline{T}}(\omega))G^{\overline{T}}(\omega)
-\wSigma^>(\omega) G^{{<}} (\omega) &=& i\mu (G^>(\omega) + G^<(\omega)). \nn\\
\ee
The combination of the previous equations results in
\be
 \wSigma^<(\omega) G^> (\omega)
-\wSigma^>(\omega) G^{{<}} (\omega) &=& i\mu (G^>(\omega) + G^<(\omega)),
\ee
which can be rewritten as
\be
\frac 12( \wSigma^<(\omega) -\wSigma^>(\omega)) (  G^< (\omega)+ G^> (\omega))
-\frac12( \wSigma^<(\omega) +\wSigma^>(\omega)) (  G^< (\omega)- G^> (\omega))
&=& i\mu (G^>(\omega) + G^<(\omega)).\nn\\
\ee
After substituting  the equations for the self-energy, we obtain one equation
with two unknown functions, resulting in  a  relation between  $G^>(\omega)$
and $G^<(\omega)$. We expect at most a finite set of solutions. If we make the
Ansatz $G^>(\omega) = \alpha \exp(b\omega) G^<(\omega) $ we find
\be
\alpha e^{b\omega}(1-\alpha^2) \Sigma^{<}(\omega) G^<(\omega)
  =i\mu (1+\alpha e^{b\omega})G^<(\omega).
\ee
Since both $G^<(\omega) $ and $\Sigma^<(\omega)$ are non-vanishing, not constant
functions,
we conclude that $\alpha=-1$ and $b=0$.

We thus have shown that the  SD equations imply
\be
G^>(\omega) = -G^<(\omega).
\label{gls}
\ee
Using the usual finite temperature relation $ G^>(\omega) = - \exp(\beta \omega) G^<(\omega)$, we confirm the solutions of the SD equations give Green's functions at $\beta=0$
in agreement with the equilibrium solution of the Lindblad equation equal to the identity,
or in vectorized form, the TFD state at $\beta=0$.
After substituting \eref{gls} into \eref{a1} we find
\be
- (\Sigma^T(\omega)+\Sigma^T(\omega))G^<(\omega)
+ \wSigma^<(\omega) (G^{\overline{T}} (\omega) + G^T(\omega))
=-i\mu  (G^{\overline{T}} (\omega) +G^{{\oT}}(\omega)).
\ee
This shows that
\be
G^\oT(\omega) = -G^T(\omega).
\ee

\section{Derivation of the OTOC kernel equation}\label{app:kernel}

In this appendix, we derive an expression for the OTOC in terms of Green's functions.
The derivation is along the lines of Ref.~\cite{maldacena2016} and follows  the
calculation of the OTOC  \cite{nosaka2021} for the Maldacena-Qi model \cite{maldacena2018}.

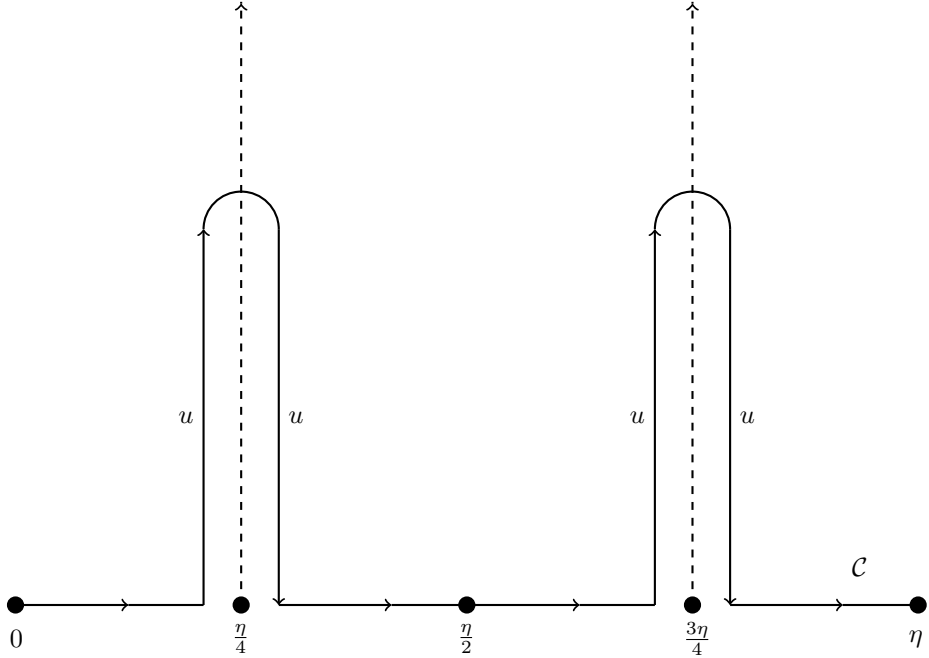
\begin{figure}[htbp]
\centering
\begin{tikzpicture}[smooth,scale=1]
\draw[thick,->] (0,0) -- (1.5,0) node()[left]{};
\draw[thick] (1.5,0) -- (2.5,0) node()[left]{};
\draw[thick,fill=black] (0,0) circle (0.1) node()[below]{$$}; 
\draw[thick,fill=black] (3,0) circle (0.1) node()[below]{$$}; 
\draw[thick,fill=black] (0,0) circle (0.0) node()[above]{}; 
\draw[thick,->] (2.5,0) -- (2.5,5) node()[left]{};
\draw[thick,->] (3.5,5) -- (3.5,0) node()[left]{};
\draw[thick] (3.5,5) arc(0:180:0.5);
\draw[thick] (2.5,2.5)  node()[left]{$u$}; 
\draw[thick] (3.5,2.5)  node()[right]{$u$}; 
\draw[thick,fill=black] (0,-5) circle (0.0) node()[above]{}; 
\draw[thick,->] (3.5,0) -- (5,0) node()[left]{};
\draw[thick] (5,0) -- (6,0) node()[left]{};
\draw[thick,fill=black] (6,0) circle (0.1) node()[below]{$$}; 
 \draw[thick,->] (6,0) -- (7.5,0) node()[left]{};
\draw[thick] (7.5,0) -- (8.5,0) node()[left]{};
\draw[thick,fill=black] (9,0) circle (0.1) node()[below]{$$}; 
\draw[thick,fill=black] (0,-5) circle (0.0) node()[above]{}; 
\draw[thick,->] (8.5,0) -- (8.5,5) node()[left]{};
\draw[thick,->] (9.5,5) -- (9.5,0) node()[left]{};
\draw[thick] (9.5,5) arc(0:180:0.5);
\draw[thick] (8.5,2.5)  node()[left]{$u$}; 
\draw[thick] (9.5,2.5)  node()[right]{$u$}; 
\draw[thick,fill=black] (0,0) circle (0.0) node()[above]{}; 
\draw[thick,->] (9.5,0) -- (11,0) node()[left]{};
\draw[thick] (11,0) -- (12,0) node()[left]{};
\draw[thick,fill=black] (12,0) circle (0.1) node()[below]{$$}; 
\node[anchor=west] at (11,0.5) {$\sC$};
\node[anchor=west] at (-0.2,-0.45) {$0$};
\node[anchor=west] at (2.75,-0.45) {$\frac \eta 4$};
\node[anchor=west] at (5.75,-0.45) {$\frac \eta 2$};
\node[anchor=west] at (8.75,-0.45) {$\frac {3\eta} 4$};
\node[anchor=west] at (11.75,-0.45) {$\eta$};
\draw[thick,->] (3,8) -- (3,8) node()[left]{};
\draw[thick,dashed] (3,0) -- (3,8) node()[left]{};
\node[anchor=west] at (3,8) {$ $};
\draw[thick,->] (9,8) -- (9,8) node()[left]{};
\draw[thick,dashed] (9,0) -- (9,8) node()[left]{};
\node[anchor=west] at (9,8) {$ $};
\end{tikzpicture}
\vspace*{-4cm}
\caption{In order to calculate the OTOC for the SYK model with dissipation,
  we introduce the Keldysh contour $\sC$ (solid line). The solid back disks and
  the dashed lines denote the values of the arguments for which the OTOC is
  evaluated.}
  \label{contour}
\end{figure}

The regularized OTOC \cite{maldacena2016,maldacena2018} is defined by
the four-point function
\be
F(t_1,t_3)=\frac {1}{N^2} \frac 1Z \left \langle \Tr  \rho^{1/4}_{\eta} \psi_i^+(t_1)
\rho^{1/4}_{\eta}\psi_j^-(0) \rho^{1/4}_{\eta}\psi_i^+(t_3)\rho^{1/4}_{\eta} \psi_j^-(0)\right \rangle,
\ee
where $\psi_i^\pm(t)$ refers to Majoranas living on the forward and backward in time sections of the contour Fig.~\ref{contour} to be introduced shortly,  $\langle \cdots \rangle$ denote averaging over the random
couplings and $\rho_\eta$ 
is a regulator for a proper definition of the Keldysh contour representing a "thermal" density matrix at inverse temperature $\eta \ll 1$, see Fig.~\ref{contour}. We will take the limit $\eta \to 0$ at the end of the calculation. The sum over repeated indexes $i, j$ is assumed and the average is
  normalized by the ''partition'' function $Z=\Tr \rho_\eta$, which is known as the dissipative form factor \cite{khramtsov2021}. 

In order to proceed, it is convenient to introduce the Euclidean 
four-point function given by
\be
F^E(u_1,u_2,u_3,u_4) &&= \frac 1{N^2}\left \langle \Tr 
  \psi_i(u_1) \psi_j(u_2)  \psi_i(u_3) \psi_j(u_4) \right \rangle,
\label{euclid}
\ee
where the expectation value is with respect to the path integral over Majorana fermions
discussed below and the Euclidean coordinates given by
\be
u_1 =it_1+\frac \eta 4, \quad u_2 =\frac \eta 2,\quad   u_3 =it_3+\frac {3\eta} 4,\quad
u_4=\eta,
\label{uk}
\ee
are time ordered on the Keldysh contour of Fig. \ref{contour}.
In this appendix
we use the Euclidean variable $u$ related to the contour integration variable
$z$ in the main text by $u = iz$. 
Below we will omit the redundant superscripts of the $\psi_i$.
The four-point function can be
expressed as a two-replica path
integral that can be evaluated by
a saddle-point approximation. 
To leading order in $1/N$ the result  factorizes into the product of
two  Green's functions,  but these do not contribute to the exponentially increasing
part of the OTOC and we will
ignore them. The exponentially increasing part of the  OTOC, ${\cal F}(t_1,0,t_3,0) $, is obtained from
the connected $1/N$ corrections\footnote{There are also disconnected $1/N$ corrections
from the contraction of $\psi_i (u_1)$ with  $\psi_j (u_4)$ and
$\psi_i (u_2)$ with  $\psi_j (u_3)$.}
\be
F(t_1,0,t_3,0) = - \frac 1{N^2}\langle  \psi_i(u_1) \psi_i(u_3)\rangle
\langle  \psi_j(u_2) \psi_j(u_4)\rangle + \frac 1N  {\cal F}  (u_1,u_2,u_3,u_4),
\ee
which is the average of the  product of two fermion bilinears $\sum_k \psi_k(u) \psi_k(v)$.

The four-point function is given by the expectation value Eq.~\eref{euclid}
  with respect to the action
\be
    -S &=& \int_\sC \left( -\frac{i}{2}\sum\psi_i(u)\partial_u \psi_i(u)
    - i^{q/2+1}\sum J_{ijkl}\psi_i(u) \psi_j(u) \psi_k(u) \psi_l(u) 
\right )\nn \\ &&
    +\frac{\mu}2\int\!\!\!\int_\sC\sum_i \psi_i(u_a)\psi_i(u_b) K(u_a,u_b) ,
\ee
where $\sC$ is the Keldysh contour shown in Fig. \ref{contour},
and $K(u,v)=-K(v,u)$ is defined by
\be
\frac 12 \int\!\!\!\int_\sC\sum_i \psi_i(u_a)\psi_i(u_b) K(u_a,u_b)
=\int_{0}^\infty dt \psi^+(t)\psi^-(t).
\ee
The limit of integration of the integration contour $\sC$ and
the right hand side of the above equation can be extended to $-\infty$ without affecting the result. We note that to simplify the notation in this appendix,
we are not writing down the differentials $du$ for single integrals,   $du_a du_b$ for double integrals, and $du_adu_b du_cdu_d$
for quadruple integrals. 

The corresponding partition function after averaging over the random couplings
is given by
\begin{equation}\begin{aligned}
    \langle Z \rangle_J &=
     \int D\psi \exp\left(-\frac i2\int_\sC \psi_i(u)\partial_u \psi_i(u)
    - \frac{N}{2q}J^2 \int\!\!\!\int_\sC \left(\frac{1}{N} \psi_{i}(u_a)\psi_{i}(u_b)\right)^q + \frac \mu 2
    \int_\sC  \psi_i(u_a) \psi_i(u_b)K(u_a,u_b)  \right). 
\end{aligned}\end{equation}
Introducing the  identity 
\begin{equation}\begin{aligned}
1 \sim \int DG D\Sigma \exp\left( -\frac{N}{2}\int\!\!\!\int_\sC \Sigma(u_a,u_b)\left(G(u_a,u_b)+\frac{i}{N}\sum_i\psi_i(u_a)\psi_i(u_b)\right) \right)
\end{aligned}\end{equation}
the partition function becomes (note that $q$ is even)
\be
\langle Z \rangle_J 
&=&\int DG D\Sigma D\psi \exp\left(-\frac{i}{2}\int_\sC \psi_i(u)\partial_u\psi_i(u) - \frac{N}{2q}J^2 \int\!\!\!\int_\sC \left[i G(u_a,u_b)\right]^q \right.\\
&&\left. - \frac{iN\mu}{2} \int\!\!\!\int_\sC  G(u_a,u_b)K(u_a,u_b)  -\frac{N}{2}\int\!\!\!\int_\sC \Sigma(u_a,u_b) G(u_a,u_b) -\frac{i}{2} \int\!\!\!\int_\sC \Sigma(u_a,u_b)\psi_i(u_a)\psi_i(u_b)  \right) . \nn
\ee
  After integrating over the fermions, we obtain the
  $\Sigma G$ expression  for the partition function,
\begin{equation}\begin{aligned}
    \langle Z\rangle_J &= \int DG D\Sigma \exp \left( \frac N2 \int\int_\sC
    \Tr \log\left(i\delta(u_a,u_b)\partial_{u_a} + i\Sigma(u_a,u_b) \right)
     -\frac{N}{2q}J^2 \int\!\!\!\int_\sC \left[i G(u_a,u_b)\right]^q \right ) \\
    &\times  \exp\left( +\frac{iN\mu}{2} \int\!\!\!\int_\sC G(u_a,u_b)K(u_a,u_b)
    -\frac{N}{2}\int\!\!\!\int_\sC \Sigma(u_a,u_b) G(u_a,u_b)\right).  \\
\end{aligned}\end{equation}
Variation with respect to $\delta G(u_a,u_b)$ and  $\delta \Sigma(u_a,u_b)$
gives the Schwinger-Dyson equations
\begin{equation}\begin{aligned}
   - \Sigma(u_a,u_b) - i^q J^2 G(u_a,u_b)^{q-1} + i\mu K(u_a,u_b)\delta(u_a,u_b) = 0, \nn\\
    (-\partial_u -\Sigma)\cdot  G = 1.
    \label{eq:SD_Sigma}
\end{aligned}\end{equation}
In terms of the $z$ variables along the Keldysh contour of the main text, the second
equation becomes
\be
    (i\partial_z -\Sigma)\cdot  G = 1,
\ee
  while the first equation remains the same. The OTOC is given by the leading $1/N$ corrections, $\mathcal{F}(u_1,u_2,u_3,u_4)/N$, of
\be
F(t_1,t_2=0,t_3,t_4=0) = -\frac 1 Z \langle G(u_1,u_3) G(u_2,u_4) \rangle
\ee
with $u_k$ defined in Eq. \eref{uk}. We will evaluate this correlation function
for general Euclidean $u_k$. The $1/N$ corrections
 are obtained from 
the second order expansion of the action about the saddle point.
Writing $\Sigma(u,v) = \overline{\Sigma}(u,v) + \delta \Sigma(u,v)$ and
  $G(u,v) = \overline{G}(u,v) +\delta G(u,v)$ we find
\be
&&-\frac N4 \int\!\!\!\int\!\!\!\int\!\!\!\int_\sC
\bar G(u_a,u_b) \delta \Sigma(u_b,u_c) \bar G(u_c,u_d)  \delta \Sigma(u_d,u_a)
-\frac N2 \int\!\!\! \int_\sC\delta \Sigma(u_a,u_b) \delta G(u_a,u_b)
\nn\\
&&-\frac N4 J^2 (q-1)i^q \int \!\!\!\int_\sC  \oG(u_a,u_b)^{q-2}\delta G(u_a,u_b)^2.
\ee
The integral over $\delta \Sigma(u_a,u_b)$ can be performed by completing squares
resulting in the effective action
\be
S^{\rm eff}&=&\frac N4 \int\!\!\!\int\!\!\!\int\!\!\!\int_C
\bar G^{-1}(u_c,u_d)  \delta G(u_d,u_a)   \bar G^{-1}(u_a,u_b) \delta G(u_b,u_c)
\nn\\
&&-\frac N4 J^2 (q-1)i^q\int\!\!\!\int_\sC  \oG(u_a,u_b)^{q-2}\delta G(u_a,u_b)^2\nn\\
&=&
\frac N4 \int\!\!\!\int\!\!\!\int\!\!\!\int_\sC
\oG^{-1}(u_a,u_c) \oG^{-1}(u_b,u_d) 
\left [ \delta(u_a-u_f)\delta (u_b-u_e)  \right . \nn\\
  && 
  +\frac N4 J^2 (q-1)i^q \left . \oG(u_f,u_a) \oG(u_e,u_b) \oG^{q-2}(u_e,u_f)\right ] \delta G(u_e,u_c)    \delta G(u_f,u_d).
\ee
The $1/N$ correction to the four-point function
\be
{\cal F}(u_1,u_2,u_3,u_4) = -\int d [\delta G] \delta G(u_1,u_3) \delta G(u_2,u_4) e^{-S^{\rm eff}}
\ee
 is evaluated as
\be
   {\cal F}(u_1,u_2,u_3,u_4) &=& -\left [ \id - K\right]^{-1}(u_1,u_3;u,v) \oG(u,u_2) \oG(v,u_4)
    \\ &=& -\left[ \oG(u_3,u_2) \oG(u_1,u_4)  
  - J^2(q-1) i^q   \int \!\!\! \int  K(u_1,u_3,u,v)   \oG(u,u_2) \oG(v,u_4) +\cdots \right ]\nn
   \ee
   with kernel $K$ given by
   \be
   K(u_a,u_b,u_c,u_d) = -J^2(q-1) i^q \oG(u_a,u_c)  \oG(u_b,u_d) \oG^{q-2} (u_c,u_d),
   \ee
  and the points $u_k$ are equal to
   \be
u_1 =it_1+\frac \eta 4, \quad u_2 =+\frac \eta 2,\quad   u_3 =it_3+\frac {3\eta} 4,\quad
u_4=+\eta.
\ee
The action of the kernel $K$ by expanding the inverse in a
geometric series results in the combinations
\be
\int\!\!\! \int_\sC du dv K(it_1+\frac \eta 4,  it_3+\frac {3\eta}4,u,v ) F(u,v) .
\ee
As discussed in Ref.~\cite{Gu:2018jsv,nosaka2021},
the part of the $u$ integration that is on the $3\eta/4$ branch cancels because
the upward integration differs by a minus sign from the downward integration (the
$\epsilon$ difference between the upward and downward path is negligible with respect to $3\eta/4$ because we take the $\epsilon \to 0$ limit first).
This is not the case when the $u$ integration is over the $\eta/4$ branch.  Then, when $t_1$  is on the + path,
\be
i(G(t_1^+, t_u^+)- G(t_1^+, t_u^-))= i(G^T(t_1, t_u) - G^<(t_1, t_u)) =iG^R(t_1, t_u).
\ee
The result is the same when $t_1$ is on the minus path
\be
i(G(t_1^-, t_u^+)- G(t_1^-, t_u^-))=  i( G^>(t_1, t_u)-G^{\bar T}(t_1, t_u)) =iG^R(t_1, t_u).
\ee
This also means that the $u$ variable only contributes when it is on the $\eta/4$ branch.
An analogue argument shows that the $v$ integration only contributes when it is on
the $3\eta/4$
branch. Combining the $+$ and $-$ paths again gives the retarded Green's function
$
G^R(t_3, t_v).
$
In the domain where $u$ and $v$ give non-vanishing contributions, they are separated
by $\eta/2$. Therefore the Green's function $G(u,v)$
simply becomes the Wightman function
$ G(it_u ,it_v+\eta/2) \equiv  G_W(i(t_u-t_b) -\eta/2)$.

We conclude that the kernel is given by
\be
K(t_1,t_2,t_3,t_4) = -J^2(q-1)G^R(t_1,t_3) G^R(t_2,t_4) G_W^{q-2}(t_3,t_4).
\ee
In the domain of exponential growth, for large $t_1$ and $t_2$, we have that
$(\id - K) \mathcal{F} \ll \mathcal{F}_0$ with
$\mathcal{F}_0(u_1,u_2,u_3,u_4) = \oG(u_1,u_3) \oG(u_2,u_4)$.
Neglecting the right hand side, the Lyapunov exponent is
  obtained by solving the integral equation
  \be
  (\id - K) \mathcal{F} =0.
    \ee
\section{Solution of kernel equation for $\beta = 0$}\label{app:solkerb0}

In this Appendix, we calculate the integral in Eq. \eref{eq:ladder} of the main text, 
\be
I = \iint dt_3 dt_4 \theta(t_1 -t_3)\theta(t_2 -t_4)  e^{-\sigma (t_1 -t_3)}
e^{-\sigma (t_2 -t_4)} \frac{2\alpha^2}{ \cosh^2(\alpha|t_3 -t_4|+\gamma)}f(|t_3-t_4)|),
\label{integral}
\ee
  where $f(|t_3-t_4)|)$ is the solution of the differential equation
  \eref{eq:ladder_derivative},
  \be
  (\partial_t^2 -\sigma^2)f(t) = - \frac{2\alpha^2}{ \cosh^2(\alpha|t|+\gamma)} f(t).
\label{difeq}
\ee
 We will check that the integral \eref{integral} is given by $f(t_{12})$.
  The integral can be carried out by changing coordinates as follows, 
  \be
  s =t_3- t_4 \quad {\rm and } \quad t = \frac 12 (t_3+t_4).
  \ee
  This results in
  \be
 I= \int_0^\infty dt e^{-2\sigma t}
  \int_{-2t}^{2t}ds  \frac{2\alpha^2}{ \cosh^2(\alpha|s+t_{12}|+\gamma)} f(|s+t_{12}|),
  \ee
  where $t_{12}= t_1-t_2$. After a partial integration of the $t$ integral, integrating over
  $s$ 
  results in
  \be
  I&=&
     \frac 1{2\sigma} \int_0^\infty dt  e^{-2\sigma t}
     \left( \frac{2\alpha^2}{ \cosh^2(\alpha|2t + t_{12} |+\gamma)} f(|2t+t_{12}|)
     + \frac{2\alpha^2}{ \cosh^2(\alpha|-2t + t_{12} |+\gamma)} f(|-2t+t_{12}|)
     \right )\nn\\
     &=&
  \frac 1{\sigma} \int_{t_{12}}^\infty du  e^{-\sigma (u-t_{12})}
    \frac{2\alpha^2}{ \cosh^2(\alpha|u |+\gamma)} f(|u|)
     +
\frac 1{\sigma} \int^{t_{12}}_{-\infty} du  e^{\sigma (u-t_{12})}
     \frac{2\alpha^2}{ \cosh^2(\alpha |u| +\gamma)} f(|u|).
     \ee
     Next we use that $f(u)$ satisfies the differential equation \eref{difeq} resulting in
     \be
    I &=&
    \frac 1{2\sigma} \int_{t_{12}}^\infty du  e^{-\sigma (u-t_{12})}
    (\sigma^2-\del_u^2) f(|u|)
     +
\frac 1{2\sigma} \int^{t_{12}}_{0} du  e^{\sigma (u-t_{12})}
(\sigma^2-\del_u^2) f(|u|)\nn\\
    &&+
\frac 1{2\sigma} \int^{0}_{-\infty} du  e^{\sigma (u-t_{12})}
(\sigma^2-\del_u^2) f(|u|).
     \ee    
     The $\sigma^2 $ term cancels against the integral that remains after two partial
     integrations of the $\del_u^2$ term leaving us with only boundary contributions
     given by
     \be
     I &=& -\frac1{2\sigma} \left. e^{-\sigma (u-t_{12})} \del_u f(|u|)\right|_{t_{12}}^\infty
     -\frac 12
          \left.  e^{-\sigma (u-t_{12})} f(|u|)\right|_{t_{12}}^\infty \nn\\ &&
 - \frac1{2\sigma}\left. e^{\sigma (u-t_{12})} \del_u f(|u|)\right|^{t_{12}}_{0_+}
            - \frac 12\left.  e^{\sigma (u-t_{12})}f(|u|)\right|^{t_{12}}_{0_+}
\nn \\ && - \frac1{2\sigma}\left. e^{-\sigma t_{12}} \del_u f(|u|)\right|_{-\infty}^{0_-}
            + \frac 12\left.  e^{-\sigma t_{12}}f(|u|)\right|_{-\infty}^{0_-}.
            \ee
            The solution $f(|u|)$ is a function of $|u|$ so that the derivatives at $0_-$ and
            at $0_+$ have the opposite sign. We finally obtain
            \be
            I =  f(t_{12}) + \frac1{2\sigma}\left. e^{\sigma (-t_{12})} \del_u f(|u|)\right|_{0_-}^{0_+}.
            \ee
            Therefore, the integral $I = f(t_{12})$ if  $f'(u=0) =0$, which
            is the same condition as the one obtained from solving the differential equation
            in the main text.

\bibliography{libotoclind3}

\begin{thebibliography}{97}%
\makeatletter
\providecommand \@ifxundefined [1]{%
 \@ifx{#1\undefined}
}%
\providecommand \@ifnum [1]{%
 \ifnum #1\expandafter \@firstoftwo
 \else \expandafter \@secondoftwo
 \fi
}%
\providecommand \@ifx [1]{%
 \ifx #1\expandafter \@firstoftwo
 \else \expandafter \@secondoftwo
 \fi
}%
\providecommand \natexlab [1]{#1}%
\providecommand \enquote  [1]{``#1''}%
\providecommand \bibnamefont  [1]{#1}%
\providecommand \bibfnamefont [1]{#1}%
\providecommand \citenamefont [1]{#1}%
\providecommand \href@noop [0]{\@secondoftwo}%
\providecommand \href [0]{\begingroup \@sanitize@url \@href}%
\providecommand \@href[1]{\@@startlink{#1}\@@href}%
\providecommand \@@href[1]{\endgroup#1\@@endlink}%
\providecommand \@sanitize@url [0]{\catcode `\\12\catcode `\$12\catcode
  `\&12\catcode `\#12\catcode `\^12\catcode `\_12\catcode `\%12\relax}%
\providecommand \@@startlink[1]{}%
\providecommand \@@endlink[0]{}%
\providecommand \url  [0]{\begingroup\@sanitize@url \@url }%
\providecommand \@url [1]{\endgroup\@href {#1}{\urlprefix }}%
\providecommand \urlprefix  [0]{URL }%
\providecommand \Eprint [0]{\href }%
\providecommand \doibase [0]{https://doi.org/}%
\providecommand \selectlanguage [0]{\@gobble}%
\providecommand \bibinfo  [0]{\@secondoftwo}%
\providecommand \bibfield  [0]{\@secondoftwo}%
\providecommand \translation [1]{[#1]}%
\providecommand \BibitemOpen [0]{}%
\providecommand \bibitemStop [0]{}%
\providecommand \bibitemNoStop [0]{.\EOS\space}%
\providecommand \EOS [0]{\spacefactor3000\relax}%
\providecommand \BibitemShut  [1]{\csname bibitem#1\endcsname}%
\let\auto@bib@innerbib\@empty
\bibitem [{\citenamefont {Sekino}\ and\ \citenamefont
  {Susskind}(2008)}]{sekino2008}%
  \BibitemOpen
  \bibfield  {author} {\bibinfo {author} {\bibfnamefont {Y.}~\bibnamefont
  {Sekino}}\ and\ \bibinfo {author} {\bibfnamefont {L.}~\bibnamefont
  {Susskind}},\ }\bibfield  {title} {\bibinfo {title} {Fast scramblers},\
  }\href@noop {} {\bibfield  {journal} {\bibinfo  {journal} {Journal of High
  Energy Physics}\ }\textbf {\bibinfo {volume} {10}},\ \bibinfo {pages} {065}
  (\bibinfo {year} {2008})}\BibitemShut {NoStop}%
\bibitem [{\citenamefont {Bagrets}\ \emph {et~al.}(2017)\citenamefont
  {Bagrets}, \citenamefont {Altland},\ and\ \citenamefont
  {Kamenev}}]{bagrets2017}%
  \BibitemOpen
  \bibfield  {author} {\bibinfo {author} {\bibfnamefont {D.}~\bibnamefont
  {Bagrets}}, \bibinfo {author} {\bibfnamefont {A.}~\bibnamefont {Altland}},\
  and\ \bibinfo {author} {\bibfnamefont {A.}~\bibnamefont {Kamenev}},\
  }\bibfield  {title} {\bibinfo {title} {Power-law out of time order
  correlation functions in the {SYK} model},\ }\href
  {https://doi.org/https://doi.org/10.1016/j.nuclphysb.2017.06.012} {\bibfield
  {journal} {\bibinfo  {journal} {Nuclear Physics B}\ }\textbf {\bibinfo
  {volume} {921}},\ \bibinfo {pages} {727 } (\bibinfo {year}
  {2017})}\BibitemShut {NoStop}%
\bibitem [{\citenamefont {Larkin}\ and\ \citenamefont
  {Ovchinnikov}(1969)}]{larkin1969}%
  \BibitemOpen
  \bibfield  {author} {\bibinfo {author} {\bibfnamefont {A.}~\bibnamefont
  {Larkin}}\ and\ \bibinfo {author} {\bibfnamefont {Y.~N.}\ \bibnamefont
  {Ovchinnikov}},\ }\bibfield  {title} {\bibinfo {title} {Quasiclassical method
  in the theory of superconductivity},\ }\href@noop {} {\bibfield  {journal}
  {\bibinfo  {journal} {Sov Phys JETP}\ }\textbf {\bibinfo {volume} {28}},\
  \bibinfo {pages} {1200} (\bibinfo {year} {1969})}\BibitemShut {NoStop}%
\bibitem [{\citenamefont {Berman}\ and\ \citenamefont
  {Zaslavsky}(1978)}]{berman1978}%
  \BibitemOpen
  \bibfield  {author} {\bibinfo {author} {\bibfnamefont {G.}~\bibnamefont
  {Berman}}\ and\ \bibinfo {author} {\bibfnamefont {G.}~\bibnamefont
  {Zaslavsky}},\ }\bibfield  {title} {\bibinfo {title} {Condition of
  stochasticity in quantum nonlinear systems},\ }\href
  {https://doi.org/http://dx.doi.org/10.1016/0378-4371(78)90190-5} {\bibfield
  {journal} {\bibinfo  {journal} {Physica A: Statistical Mechanics and its
  Applications}\ }\textbf {\bibinfo {volume} {91}},\ \bibinfo {pages} {450 }
  (\bibinfo {year} {1978})}\BibitemShut {NoStop}%
\bibitem [{\citenamefont {Bohigas}\ \emph {et~al.}(1984)\citenamefont
  {Bohigas}, \citenamefont {Giannoni},\ and\ \citenamefont
  {Schmit}}]{bohigas1984}%
  \BibitemOpen
  \bibfield  {author} {\bibinfo {author} {\bibfnamefont {O.}~\bibnamefont
  {Bohigas}}, \bibinfo {author} {\bibfnamefont {M.~J.}\ \bibnamefont
  {Giannoni}},\ and\ \bibinfo {author} {\bibfnamefont {C.}~\bibnamefont
  {Schmit}},\ }\bibfield  {title} {\bibinfo {title} {{Characterization of
  Chaotic Quantum Spectra and Universality of Level Fluctuation Laws}},\ }\href
  {https://doi.org/10.1103/PhysRevLett.52.1} {\bibfield  {journal} {\bibinfo
  {journal} {Phys. Rev. Lett.}\ }\textbf {\bibinfo {volume} {52}},\ \bibinfo
  {pages} {1} (\bibinfo {year} {1984})}\BibitemShut {NoStop}%
\bibitem [{\citenamefont {Wigner}(1951)}]{wigner1951}%
  \BibitemOpen
  \bibfield  {author} {\bibinfo {author} {\bibfnamefont {E.}~\bibnamefont
  {Wigner}},\ }\bibfield  {title} {\bibinfo {title} {On the statistical
  distribution of the widths and spacings of nuclear resonance levels},\ }\href
  {https://doi.org/10.1017/S0305004100027237} {\bibfield  {journal} {\bibinfo
  {journal} {Math. Proc. Cam. Phil. Soc.}\ }\textbf {\bibinfo {volume} {49}},\
  \bibinfo {pages} {790} (\bibinfo {year} {1951})}\BibitemShut {NoStop}%
\bibitem [{\citenamefont {Dyson}(1962{\natexlab{a}})}]{dyson1962a}%
  \BibitemOpen
  \bibfield  {author} {\bibinfo {author} {\bibfnamefont {F.~J.}\ \bibnamefont
  {Dyson}},\ }\bibfield  {title} {\bibinfo {title} {{Statistical theory of the
  energy levels of complex systems. I}},\ }\href
  {https://doi.org/10.1063/1.1703773} {\bibfield  {journal} {\bibinfo
  {journal} {J. Math. Phys.}\ }\textbf {\bibinfo {volume} {3}},\ \bibinfo
  {pages} {140} (\bibinfo {year} {1962}{\natexlab{a}})}\BibitemShut {NoStop}%
\bibitem [{\citenamefont {Dyson}(1962{\natexlab{b}})}]{dyson1962b}%
  \BibitemOpen
  \bibfield  {author} {\bibinfo {author} {\bibfnamefont {F.}~\bibnamefont
  {Dyson}},\ }\bibfield  {title} {\bibinfo {title} {{Statistical Theory of the
  Energy Levels of Complex Systems. II}},\ }\href
  {https://doi.org/10.1063/1.1703774} {\bibfield  {journal} {\bibinfo
  {journal} {J. Math. Phys.}\ }\textbf {\bibinfo {volume} {3}},\ \bibinfo
  {pages} {157} (\bibinfo {year} {1962}{\natexlab{b}})}\BibitemShut {NoStop}%
\bibitem [{\citenamefont {Dyson}(1962{\natexlab{c}})}]{dyson1962c}%
  \BibitemOpen
  \bibfield  {author} {\bibinfo {author} {\bibfnamefont {F.}~\bibnamefont
  {Dyson}},\ }\bibfield  {title} {\bibinfo {title} {{Statistical Theory of the
  Energy Levels of Complex Systems. III}},\ }\href
  {https://doi.org/10.1063/1.1703775} {\bibfield  {journal} {\bibinfo
  {journal} {J. Math. Phys.}\ }\textbf {\bibinfo {volume} {3}},\ \bibinfo
  {pages} {166} (\bibinfo {year} {1962}{\natexlab{c}})}\BibitemShut {NoStop}%
\bibitem [{\citenamefont {Dyson}(1962{\natexlab{d}})}]{dyson1962d}%
  \BibitemOpen
  \bibfield  {author} {\bibinfo {author} {\bibfnamefont {F.}~\bibnamefont
  {Dyson}},\ }\bibfield  {title} {\bibinfo {title} {{The Threefold Way.
  Algebraic Structure of Symmetry Groups and Ensembles in Quantum Mechanics}},\
  }\href {https://doi.org/10.1063/1.1703863} {\bibfield  {journal} {\bibinfo
  {journal} {J. Math. Phys.}\ }\textbf {\bibinfo {volume} {3}},\ \bibinfo
  {pages} {1199} (\bibinfo {year} {1962}{\natexlab{d}})}\BibitemShut {NoStop}%
\bibitem [{\citenamefont {Dyson}(1972)}]{dyson1972}%
  \BibitemOpen
  \bibfield  {author} {\bibinfo {author} {\bibfnamefont {F.}~\bibnamefont
  {Dyson}},\ }\bibfield  {title} {\bibinfo {title} {{A Class of Matrix
  Ensembles}},\ }\href {https://doi.org/10.1063/1.1665857} {\bibfield
  {journal} {\bibinfo  {journal} {J. Math. Phys.}\ }\textbf {\bibinfo {volume}
  {13}},\ \bibinfo {pages} {90} (\bibinfo {year} {1972})}\BibitemShut {NoStop}%
\bibitem [{\citenamefont {Mehta}(2004)}]{mehta2004}%
  \BibitemOpen
  \bibfield  {author} {\bibinfo {author} {\bibfnamefont {M.~L.}\ \bibnamefont
  {Mehta}},\ }\href@noop {} {\emph {\bibinfo {title} {Random matrices}}}\
  (\bibinfo  {publisher} {Academic press},\ \bibinfo {year} {2004})\BibitemShut
  {NoStop}%
\bibitem [{\citenamefont {Maldacena}\ \emph {et~al.}(2016)\citenamefont
  {Maldacena}, \citenamefont {Shenker},\ and\ \citenamefont
  {Stanford}}]{maldacena2015}%
  \BibitemOpen
  \bibfield  {author} {\bibinfo {author} {\bibfnamefont {J.}~\bibnamefont
  {Maldacena}}, \bibinfo {author} {\bibfnamefont {S.~H.}\ \bibnamefont
  {Shenker}},\ and\ \bibinfo {author} {\bibfnamefont {D.}~\bibnamefont
  {Stanford}},\ }\bibfield  {title} {\bibinfo {title} {A bound on chaos},\
  }\href {https://doi.org/10.1007/JHEP08(2016)106} {\bibfield  {journal}
  {\bibinfo  {journal} {Journal of High Energy Physics}\ }\textbf {\bibinfo
  {volume} {08}},\ \bibinfo {pages} {106} (\bibinfo {year} {2016})},\ \Eprint
  {https://arxiv.org/abs/1503.01409} {arXiv:1503.01409 [hep-th]} \BibitemShut
  {NoStop}%
\bibitem [{\citenamefont {Kitaev}(2015)}]{kitaev2015}%
  \BibitemOpen
  \bibfield  {author} {\bibinfo {author} {\bibfnamefont {A.}~\bibnamefont
  {Kitaev}},\ }\href {http://online.kitp.ucsb.edu/online/entangled15/}
  {\bibinfo {title} {A simple model of quantum holography}} (\bibinfo {year}
  {2015}),\ \bibinfo {note} {string seminar at KITP and Entanglement 2015
  program, 12 February, 7 April and 27 May 2015,
  http://online.kitp.ucsb.edu/online/entangled15/}\BibitemShut {NoStop}%
\bibitem [{\citenamefont {French}\ and\ \citenamefont
  {Wong}(1970)}]{french1970}%
  \BibitemOpen
  \bibfield  {author} {\bibinfo {author} {\bibfnamefont {J.}~\bibnamefont
  {French}}\ and\ \bibinfo {author} {\bibfnamefont {S.}~\bibnamefont {Wong}},\
  }\bibfield  {title} {\bibinfo {title} {{Validity of random matrix theories
  for many-particle systems}},\ }\href
  {https://doi.org/http://dx.doi.org/10.1016/0370-2693(70)90213-3} {\bibfield
  {journal} {\bibinfo  {journal} {Physics Letters B}\ }\textbf {\bibinfo
  {volume} {33}},\ \bibinfo {pages} {449 } (\bibinfo {year}
  {1970})}\BibitemShut {NoStop}%
\bibitem [{\citenamefont {Bohigas}\ and\ \citenamefont
  {Flores}(1971{\natexlab{a}})}]{bohigas1971}%
  \BibitemOpen
  \bibfield  {author} {\bibinfo {author} {\bibfnamefont {O.}~\bibnamefont
  {Bohigas}}\ and\ \bibinfo {author} {\bibfnamefont {J.}~\bibnamefont
  {Flores}},\ }\bibfield  {title} {\bibinfo {title} {{Two-body random
  hamiltonian and level density}},\ }\href
  {https://doi.org/http://dx.doi.org/10.1016/0370-2693(71)90598-3} {\bibfield
  {journal} {\bibinfo  {journal} {Physics Letters B}\ }\textbf {\bibinfo
  {volume} {34}},\ \bibinfo {pages} {261 } (\bibinfo {year}
  {1971}{\natexlab{a}})}\BibitemShut {NoStop}%
\bibitem [{\citenamefont {French}\ and\ \citenamefont
  {Wong}(1971)}]{french1971}%
  \BibitemOpen
  \bibfield  {author} {\bibinfo {author} {\bibfnamefont {J.}~\bibnamefont
  {French}}\ and\ \bibinfo {author} {\bibfnamefont {S.}~\bibnamefont {Wong}},\
  }\bibfield  {title} {\bibinfo {title} {{Some random-matrix level and spacing
  distributions for fixed-particle-rank interactions}},\ }\href
  {https://doi.org/http://dx.doi.org/10.1016/0370-2693(71)90424-2} {\bibfield
  {journal} {\bibinfo  {journal} {Physics Letters B}\ }\textbf {\bibinfo
  {volume} {35}},\ \bibinfo {pages} {5 } (\bibinfo {year} {1971})}\BibitemShut
  {NoStop}%
\bibitem [{\citenamefont {Bohigas}\ and\ \citenamefont
  {Flores}(1971{\natexlab{b}})}]{bohigas1971a}%
  \BibitemOpen
  \bibfield  {author} {\bibinfo {author} {\bibfnamefont {O.}~\bibnamefont
  {Bohigas}}\ and\ \bibinfo {author} {\bibfnamefont {J.}~\bibnamefont
  {Flores}},\ }\bibfield  {title} {\bibinfo {title} {{Spacing and individual
  eigenvalue distributions of two-body random hamiltonians}},\ }\href
  {https://doi.org/http://dx.doi.org/10.1016/0370-2693(71)90399-6} {\bibfield
  {journal} {\bibinfo  {journal} {Physics Letters B}\ }\textbf {\bibinfo
  {volume} {35}},\ \bibinfo {pages} {383 } (\bibinfo {year}
  {1971}{\natexlab{b}})}\BibitemShut {NoStop}%
\bibitem [{\citenamefont {Benet}\ \emph {et~al.}(2001)\citenamefont {Benet},
  \citenamefont {Rupp},\ and\ \citenamefont {Weidenm\"uller}}]{benet2001}%
  \BibitemOpen
  \bibfield  {author} {\bibinfo {author} {\bibfnamefont {L.}~\bibnamefont
  {Benet}}, \bibinfo {author} {\bibfnamefont {T.}~\bibnamefont {Rupp}},\ and\
  \bibinfo {author} {\bibfnamefont {H.~A.}\ \bibnamefont {Weidenm\"uller}},\
  }\bibfield  {title} {\bibinfo {title} {Nonuniversal behavior of the
  $\mathit{k}$-body embedded gaussian unitary ensemble of random matrices},\
  }\href {https://doi.org/10.1103/PhysRevLett.87.010601} {\bibfield  {journal}
  {\bibinfo  {journal} {Phys. Rev. Lett.}\ }\textbf {\bibinfo {volume} {87}},\
  \bibinfo {pages} {010601} (\bibinfo {year} {2001})},\ \Eprint
  {https://arxiv.org/abs/cond-mat/0010425} {arXiv:cond-mat/0010425 [cond-mat]}
  \BibitemShut {NoStop}%
\bibitem [{\citenamefont {Sachdev}\ and\ \citenamefont
  {Ye}(1993)}]{sachdev1993}%
  \BibitemOpen
  \bibfield  {author} {\bibinfo {author} {\bibfnamefont {S.}~\bibnamefont
  {Sachdev}}\ and\ \bibinfo {author} {\bibfnamefont {J.}~\bibnamefont {Ye}},\
  }\bibfield  {title} {\bibinfo {title} {{Gapless spin-fluid ground state in a
  random quantum {Heisenberg} magnet}},\ }\href
  {https://doi.org/10.1103/PhysRevLett.70.3339} {\bibfield  {journal} {\bibinfo
   {journal} {Phys. Rev. Lett.}\ }\textbf {\bibinfo {volume} {70}},\ \bibinfo
  {pages} {3339} (\bibinfo {year} {1993})},\ \Eprint
  {https://arxiv.org/abs/cond-mat/9212030} {arXiv:cond-mat/9212030 [cond-mat]}
  \BibitemShut {NoStop}%
\bibitem [{\citenamefont {Sachdev}(2010)}]{sachdev2010}%
  \BibitemOpen
  \bibfield  {author} {\bibinfo {author} {\bibfnamefont {S.}~\bibnamefont
  {Sachdev}},\ }\bibfield  {title} {\bibinfo {title} {{Holographic Metals and
  the Fractionalized Fermi Liquid}},\ }\href
  {https://doi.org/10.1103/PhysRevLett.105.151602} {\bibfield  {journal}
  {\bibinfo  {journal} {Phys. Rev. Lett.}\ }\textbf {\bibinfo {volume} {105}},\
  \bibinfo {pages} {151602} (\bibinfo {year} {2010})}\BibitemShut {NoStop}%
\bibitem [{\citenamefont {Maldacena}\ and\ \citenamefont
  {Stanford}(2016)}]{maldacena2016}%
  \BibitemOpen
  \bibfield  {author} {\bibinfo {author} {\bibfnamefont {J.}~\bibnamefont
  {Maldacena}}\ and\ \bibinfo {author} {\bibfnamefont {D.}~\bibnamefont
  {Stanford}},\ }\bibfield  {title} {\bibinfo {title} {{Remarks on the
  {Sachdev-Ye-Kitaev} model}},\ }\href
  {https://doi.org/10.1103/PhysRevD.94.106002} {\bibfield  {journal} {\bibinfo
  {journal} {Phys. Rev. D}\ }\textbf {\bibinfo {volume} {94}},\ \bibinfo
  {pages} {106002} (\bibinfo {year} {2016})}\BibitemShut {NoStop}%
\bibitem [{\citenamefont {Kobrin}\ \emph {et~al.}(2021)\citenamefont {Kobrin},
  \citenamefont {Yang}, \citenamefont {Kahanamoku-Meyer}, \citenamefont
  {Olund}, \citenamefont {Moore}, \citenamefont {Stanford},\ and\ \citenamefont
  {Yao}}]{Kobrin:2020xms}%
  \BibitemOpen
  \bibfield  {author} {\bibinfo {author} {\bibfnamefont {B.}~\bibnamefont
  {Kobrin}}, \bibinfo {author} {\bibfnamefont {Z.}~\bibnamefont {Yang}},
  \bibinfo {author} {\bibfnamefont {G.~D.}\ \bibnamefont {Kahanamoku-Meyer}},
  \bibinfo {author} {\bibfnamefont {C.~T.}\ \bibnamefont {Olund}}, \bibinfo
  {author} {\bibfnamefont {J.~E.}\ \bibnamefont {Moore}}, \bibinfo {author}
  {\bibfnamefont {D.}~\bibnamefont {Stanford}},\ and\ \bibinfo {author}
  {\bibfnamefont {N.~Y.}\ \bibnamefont {Yao}},\ }\bibfield  {title} {\bibinfo
  {title} {{Many-Body Chaos in the Sachdev-Ye-Kitaev Model}},\ }\href
  {https://doi.org/10.1103/PhysRevLett.126.030602} {\bibfield  {journal}
  {\bibinfo  {journal} {Phys. Rev. Lett.}\ }\textbf {\bibinfo {volume} {126}},\
  \bibinfo {pages} {030602} (\bibinfo {year} {2021})},\ \Eprint
  {https://arxiv.org/abs/2002.05725} {arXiv:2002.05725 [hep-th]} \BibitemShut
  {NoStop}%
\bibitem [{\citenamefont {Garc\'\i{}a-Garc\'\i{}a}\ \emph
  {et~al.}(2023)\citenamefont {Garc\'\i{}a-Garc\'\i{}a}, \citenamefont {Liu},\
  and\ \citenamefont {Verbaarschot}}]{Garcia-Garcia:2023jlu}%
  \BibitemOpen
  \bibfield  {author} {\bibinfo {author} {\bibfnamefont {A.~M.}\ \bibnamefont
  {Garc\'\i{}a-Garc\'\i{}a}}, \bibinfo {author} {\bibfnamefont
  {C.}~\bibnamefont {Liu}},\ and\ \bibinfo {author} {\bibfnamefont {J.~J.~M.}\
  \bibnamefont {Verbaarschot}},\ }\href@noop {} {\bibinfo {title} {{Sparsity
  independent Lyapunov exponent in the Sachdev-Ye-Kitaev model}}} (\bibinfo
  {year} {2023}),\ \Eprint {https://arxiv.org/abs/2311.00639} {arXiv:2311.00639
  [hep-th]} \BibitemShut {NoStop}%
\bibitem [{\citenamefont {Garc\'{\i}a-Garc\'{\i}a}\ and\ \citenamefont
  {Verbaarschot}(2016)}]{garcia2016}%
  \BibitemOpen
  \bibfield  {author} {\bibinfo {author} {\bibfnamefont {A.~M.}\ \bibnamefont
  {Garc\'{\i}a-Garc\'{\i}a}}\ and\ \bibinfo {author} {\bibfnamefont {J.~J.~M.}\
  \bibnamefont {Verbaarschot}},\ }\bibfield  {title} {\bibinfo {title}
  {{Spectral and thermodynamic properties of the Sachdev-Ye-Kitaev model}},\
  }\href {https://doi.org/10.1103/PhysRevD.94.126010} {\bibfield  {journal}
  {\bibinfo  {journal} {Phys. Rev. D}\ }\textbf {\bibinfo {volume} {94}},\
  \bibinfo {pages} {126010} (\bibinfo {year} {2016})},\ \Eprint
  {https://arxiv.org/abs/1610.03816} {arXiv:1610.03816 [hep-th]} \BibitemShut
  {NoStop}%
\bibitem [{\citenamefont {Bender}\ and\ \citenamefont
  {Boettcher}(1998)}]{bender1998}%
  \BibitemOpen
  \bibfield  {author} {\bibinfo {author} {\bibfnamefont {C.~M.}\ \bibnamefont
  {Bender}}\ and\ \bibinfo {author} {\bibfnamefont {S.}~\bibnamefont
  {Boettcher}},\ }\bibfield  {title} {\bibinfo {title} {Real spectra in
  non-hermitian hamiltonians having {$PT$} symmetry},\ }\href
  {https://doi.org/10.1103/PhysRevLett.80.5243} {\bibfield  {journal} {\bibinfo
   {journal} {Phys. Rev. Lett.}\ }\textbf {\bibinfo {volume} {80}},\ \bibinfo
  {pages} {5243} (\bibinfo {year} {1998})}\BibitemShut {NoStop}%
\bibitem [{\citenamefont {Wiersig}(2014)}]{wiersig2014}%
  \BibitemOpen
  \bibfield  {author} {\bibinfo {author} {\bibfnamefont {J.}~\bibnamefont
  {Wiersig}},\ }\bibfield  {title} {\bibinfo {title} {Enhancing the sensitivity
  of frequency and energy splitting detection by using exceptional points:
  Application to microcavity sensors for single-particle detection},\ }\href
  {https://doi.org/10.1103/PhysRevLett.112.203901} {\bibfield  {journal}
  {\bibinfo  {journal} {Phys. Rev. Lett.}\ }\textbf {\bibinfo {volume} {112}},\
  \bibinfo {pages} {203901} (\bibinfo {year} {2014})}\BibitemShut {NoStop}%
\bibitem [{\citenamefont {Gu}\ \emph {et~al.}(2020)\citenamefont {Gu},
  \citenamefont {Kitaev}, \citenamefont {Sachdev},\ and\ \citenamefont
  {Tarnopolsky}}]{gu2020}%
  \BibitemOpen
  \bibfield  {author} {\bibinfo {author} {\bibfnamefont {Y.}~\bibnamefont
  {Gu}}, \bibinfo {author} {\bibfnamefont {A.}~\bibnamefont {Kitaev}}, \bibinfo
  {author} {\bibfnamefont {S.}~\bibnamefont {Sachdev}},\ and\ \bibinfo {author}
  {\bibfnamefont {G.}~\bibnamefont {Tarnopolsky}},\ }\bibfield  {title}
  {\bibinfo {title} {Notes on the complex {Sachdev-Ye-Kitaev model}},\
  }\bibfield  {journal} {\bibinfo  {journal} {Journal of High Energy Physics}\
  }\textbf {\bibinfo {volume} {2020}},\ \href
  {https://doi.org/10.1007/jhep02(2020)157} {10.1007/jhep02(2020)157} (\bibinfo
  {year} {2020})\BibitemShut {NoStop}%
\bibitem [{\citenamefont {Kawabata}\ \emph {et~al.}(2018)\citenamefont
  {Kawabata}, \citenamefont {Ashida}, \citenamefont {Katsura},\ and\
  \citenamefont {Ueda}}]{kawabata2018}%
  \BibitemOpen
  \bibfield  {author} {\bibinfo {author} {\bibfnamefont {K.}~\bibnamefont
  {Kawabata}}, \bibinfo {author} {\bibfnamefont {Y.}~\bibnamefont {Ashida}},
  \bibinfo {author} {\bibfnamefont {H.}~\bibnamefont {Katsura}},\ and\ \bibinfo
  {author} {\bibfnamefont {M.}~\bibnamefont {Ueda}},\ }\bibfield  {title}
  {\bibinfo {title} {Parity-time-symmetric topological superconductor},\ }\href
  {https://doi.org/10.1103/PhysRevB.98.085116} {\bibfield  {journal} {\bibinfo
  {journal} {Phys. Rev. B}\ }\textbf {\bibinfo {volume} {98}},\ \bibinfo
  {pages} {085116} (\bibinfo {year} {2018})}\BibitemShut {NoStop}%
\bibitem [{\citenamefont {Kawabata}\ \emph {et~al.}(2017)\citenamefont
  {Kawabata}, \citenamefont {Ashida},\ and\ \citenamefont
  {Ueda}}]{kawabata2017}%
  \BibitemOpen
  \bibfield  {author} {\bibinfo {author} {\bibfnamefont {K.}~\bibnamefont
  {Kawabata}}, \bibinfo {author} {\bibfnamefont {Y.}~\bibnamefont {Ashida}},\
  and\ \bibinfo {author} {\bibfnamefont {M.}~\bibnamefont {Ueda}},\ }\bibfield
  {title} {\bibinfo {title} {Information retrieval and criticality in
  parity-time-symmetric systems},\ }\href
  {https://doi.org/10.1103/PhysRevLett.119.190401} {\bibfield  {journal}
  {\bibinfo  {journal} {Phys. Rev. Lett.}\ }\textbf {\bibinfo {volume} {119}},\
  \bibinfo {pages} {190401} (\bibinfo {year} {2017})}\BibitemShut {NoStop}%
\bibitem [{\citenamefont {Roccati}\ \emph {et~al.}(2024)\citenamefont
  {Roccati}, \citenamefont {Balducci}, \citenamefont {Shir},\ and\
  \citenamefont {Chenu}}]{Roccati2024}%
  \BibitemOpen
  \bibfield  {author} {\bibinfo {author} {\bibfnamefont {F.}~\bibnamefont
  {Roccati}}, \bibinfo {author} {\bibfnamefont {F.}~\bibnamefont {Balducci}},
  \bibinfo {author} {\bibfnamefont {R.}~\bibnamefont {Shir}},\ and\ \bibinfo
  {author} {\bibfnamefont {A.}~\bibnamefont {Chenu}},\ }\bibfield  {title}
  {\bibinfo {title} {Diagnosing non-hermitian many-body localization and
  quantum chaos via singular value decomposition},\ }\href
  {https://doi.org/10.1103/PhysRevB.109.L140201} {\bibfield  {journal}
  {\bibinfo  {journal} {Phys. Rev. B}\ }\textbf {\bibinfo {volume} {109}},\
  \bibinfo {pages} {L140201} (\bibinfo {year} {2024})}\BibitemShut {NoStop}%
\bibitem [{\citenamefont {Li}\ \emph {et~al.}(2019)\citenamefont {Li},
  \citenamefont {Harter}, \citenamefont {Liu}, \citenamefont {de~Melo},
  \citenamefont {Joglekar},\ and\ \citenamefont {Luo}}]{li2019}%
  \BibitemOpen
  \bibfield  {author} {\bibinfo {author} {\bibfnamefont {J.}~\bibnamefont
  {Li}}, \bibinfo {author} {\bibfnamefont {A.~K.}\ \bibnamefont {Harter}},
  \bibinfo {author} {\bibfnamefont {J.}~\bibnamefont {Liu}}, \bibinfo {author}
  {\bibfnamefont {L.}~\bibnamefont {de~Melo}}, \bibinfo {author} {\bibfnamefont
  {Y.~N.}\ \bibnamefont {Joglekar}},\ and\ \bibinfo {author} {\bibfnamefont
  {L.}~\bibnamefont {Luo}},\ }\bibfield  {title} {\bibinfo {title} {Observation
  of parity-time symmetry breaking transitions in a dissipative floquet system
  of ultracold atoms},\ }\bibfield  {journal} {\bibinfo  {journal} {Nature
  Communications}\ }\textbf {\bibinfo {volume} {10}},\ \href
  {https://doi.org/10.1038/s41467-019-08596-1} {10.1038/s41467-019-08596-1}
  (\bibinfo {year} {2019})\BibitemShut {NoStop}%
\bibitem [{\citenamefont {Garc\'{\i}a-Garc\'{\i}a}\ and\ \citenamefont
  {Godet}(2021)}]{garcia2021}%
  \BibitemOpen
  \bibfield  {author} {\bibinfo {author} {\bibfnamefont {A.~M.}\ \bibnamefont
  {Garc\'{\i}a-Garc\'{\i}a}}\ and\ \bibinfo {author} {\bibfnamefont
  {V.}~\bibnamefont {Godet}},\ }\bibfield  {title} {\bibinfo {title}
  {{Euclidean wormhole in the Sachdev-Ye-Kitaev model}},\ }\href
  {https://doi.org/10.1103/PhysRevD.103.046014} {\bibfield  {journal} {\bibinfo
   {journal} {Phys. Rev. D}\ }\textbf {\bibinfo {volume} {103}},\ \bibinfo
  {pages} {046014} (\bibinfo {year} {2021})},\ \Eprint
  {https://arxiv.org/abs/2010.11633} {arXiv:2010.11633 [hep-th]} \BibitemShut
  {NoStop}%
\bibitem [{\citenamefont {Ashida}\ \emph {et~al.}(2020)\citenamefont {Ashida},
  \citenamefont {Gong},\ and\ \citenamefont {Ueda}}]{ashida2020}%
  \BibitemOpen
  \bibfield  {author} {\bibinfo {author} {\bibfnamefont {Y.}~\bibnamefont
  {Ashida}}, \bibinfo {author} {\bibfnamefont {Z.}~\bibnamefont {Gong}},\ and\
  \bibinfo {author} {\bibfnamefont {M.}~\bibnamefont {Ueda}},\ }\bibfield
  {title} {\bibinfo {title} {Non-hermitian physics},\ }\href
  {https://doi.org/10.1080/00018732.2021.1876991} {\bibfield  {journal}
  {\bibinfo  {journal} {Advances in Physics}\ }\textbf {\bibinfo {volume}
  {69}},\ \bibinfo {pages} {249} (\bibinfo {year} {2020})}\BibitemShut
  {NoStop}%
\bibitem [{\citenamefont {Grobe}\ \emph {et~al.}(1988)\citenamefont {Grobe},
  \citenamefont {Haake},\ and\ \citenamefont {Sommers}}]{grobe1988}%
  \BibitemOpen
  \bibfield  {author} {\bibinfo {author} {\bibfnamefont {R.}~\bibnamefont
  {Grobe}}, \bibinfo {author} {\bibfnamefont {F.}~\bibnamefont {Haake}},\ and\
  \bibinfo {author} {\bibfnamefont {H.-J.}\ \bibnamefont {Sommers}},\
  }\bibfield  {title} {\bibinfo {title} {{Quantum Distinction of Regular and
  Chaotic Dissipative Motion}},\ }\href
  {https://doi.org/10.1103/PhysRevLett.61.1899} {\bibfield  {journal} {\bibinfo
   {journal} {Phys. Rev. Lett.}\ }\textbf {\bibinfo {volume} {61}},\ \bibinfo
  {pages} {1899} (\bibinfo {year} {1988})}\BibitemShut {NoStop}%
\bibitem [{\citenamefont {Fyodorov}\ \emph {et~al.}(1997)\citenamefont
  {Fyodorov}, \citenamefont {Khoruzhenko},\ and\ \citenamefont
  {Sommers}}]{fyodorov1997}%
  \BibitemOpen
  \bibfield  {author} {\bibinfo {author} {\bibfnamefont {Y.~V.}\ \bibnamefont
  {Fyodorov}}, \bibinfo {author} {\bibfnamefont {B.~A.}\ \bibnamefont
  {Khoruzhenko}},\ and\ \bibinfo {author} {\bibfnamefont {H.-J.}\ \bibnamefont
  {Sommers}},\ }\bibfield  {title} {\bibinfo {title} {{Almost Hermitian Random
  Matrices: Crossover from Wigner-Dyson to Ginibre Eigenvalue Statistics}},\
  }\href {https://doi.org/10.1103/physrevlett.79.557} {\bibfield  {journal}
  {\bibinfo  {journal} {Physical Review Letters}\ }\textbf {\bibinfo {volume}
  {79}},\ \bibinfo {pages} {557} (\bibinfo {year} {1997})},\ \Eprint
  {https://arxiv.org/abs/cond-mat/9703152} {cond-mat/9703152} \BibitemShut
  {NoStop}%
\bibitem [{\citenamefont {S{\'a}}\ \emph {et~al.}(2020)\citenamefont {S{\'a}},
  \citenamefont {Ribeiro},\ and\ \citenamefont {Prosen}}]{sa2019}%
  \BibitemOpen
  \bibfield  {author} {\bibinfo {author} {\bibfnamefont {L.}~\bibnamefont
  {S{\'a}}}, \bibinfo {author} {\bibfnamefont {P.}~\bibnamefont {Ribeiro}},\
  and\ \bibinfo {author} {\bibfnamefont {T.}~\bibnamefont {Prosen}},\
  }\bibfield  {title} {\bibinfo {title} {{Spectral and steady-state properties
  of random Liouvillians}},\ }\href
  {https://doi.org/https://doi.org/10.1088/1751-8121/ab9337} {\bibfield
  {journal} {\bibinfo  {journal} {J. Phys. A: Math. Theor.}\ }\textbf {\bibinfo
  {volume} {53}},\ \bibinfo {pages} {305303} (\bibinfo {year}
  {2020})}\BibitemShut {NoStop}%
\bibitem [{\citenamefont {S\'a}\ \emph {et~al.}(2020)\citenamefont {S\'a},
  \citenamefont {Ribeiro}, \citenamefont {Can},\ and\ \citenamefont
  {Prosen}}]{sa2020}%
  \BibitemOpen
  \bibfield  {author} {\bibinfo {author} {\bibfnamefont {L.}~\bibnamefont
  {S\'a}}, \bibinfo {author} {\bibfnamefont {P.}~\bibnamefont {Ribeiro}},
  \bibinfo {author} {\bibfnamefont {T.}~\bibnamefont {Can}},\ and\ \bibinfo
  {author} {\bibfnamefont {T.}~\bibnamefont {Prosen}},\ }\bibfield  {title}
  {\bibinfo {title} {{Spectral transitions and universal steady states in
  random Kraus maps and circuits}},\ }\href
  {https://doi.org/10.1103/PhysRevB.102.134310} {\bibfield  {journal} {\bibinfo
   {journal} {Phys. Rev. B}\ }\textbf {\bibinfo {volume} {102}},\ \bibinfo
  {pages} {134310} (\bibinfo {year} {2020})}\BibitemShut {NoStop}%
\bibitem [{\citenamefont {Rubio-Garc{\'{\i}}a}\ \emph
  {et~al.}(2022)\citenamefont {Rubio-Garc{\'{\i}}a}, \citenamefont {Molina},\
  and\ \citenamefont {Dukelsky}}]{rubio2022}%
  \BibitemOpen
  \bibfield  {author} {\bibinfo {author} {\bibfnamefont {A.}~\bibnamefont
  {Rubio-Garc{\'{\i}}a}}, \bibinfo {author} {\bibfnamefont {R.}~\bibnamefont
  {Molina}},\ and\ \bibinfo {author} {\bibfnamefont {J.}~\bibnamefont
  {Dukelsky}},\ }\bibfield  {title} {\bibinfo {title} {From integrability to
  chaos in quantum liouvillians},\ }\bibfield  {journal} {\bibinfo  {journal}
  {{SciPost} Physics Core}\ }\textbf {\bibinfo {volume} {5}},\ \href
  {https://doi.org/10.21468/scipostphyscore.5.2.026}
  {10.21468/scipostphyscore.5.2.026} (\bibinfo {year} {2022})\BibitemShut
  {NoStop}%
\bibitem [{\citenamefont {Garc\'{\i}a-Garc\'{\i}a}\ \emph
  {et~al.}(2022{\natexlab{a}})\citenamefont {Garc\'{\i}a-Garc\'{\i}a},
  \citenamefont {S\'a},\ and\ \citenamefont {Verbaarschot}}]{garcia2022d}%
  \BibitemOpen
  \bibfield  {author} {\bibinfo {author} {\bibfnamefont {A.~M.}\ \bibnamefont
  {Garc\'{\i}a-Garc\'{\i}a}}, \bibinfo {author} {\bibfnamefont
  {L.}~\bibnamefont {S\'a}},\ and\ \bibinfo {author} {\bibfnamefont {J.~J.~M.}\
  \bibnamefont {Verbaarschot}},\ }\bibfield  {title} {\bibinfo {title}
  {{Symmetry Classification and Universality in Non-Hermitian Many-Body Quantum
  Chaos by the {Sachdev-Ye-Kitaev Model}}},\ }\href
  {https://doi.org/10.1103/PhysRevX.12.021040} {\bibfield  {journal} {\bibinfo
  {journal} {Phys. Rev. X}\ }\textbf {\bibinfo {volume} {12}},\ \bibinfo
  {pages} {021040} (\bibinfo {year} {2022}{\natexlab{a}})}\BibitemShut
  {NoStop}%
\bibitem [{\citenamefont {Li}\ \emph {et~al.}(2021)\citenamefont {Li},
  \citenamefont {Prosen},\ and\ \citenamefont {Chan}}]{li2021a}%
  \BibitemOpen
  \bibfield  {author} {\bibinfo {author} {\bibfnamefont {J.}~\bibnamefont
  {Li}}, \bibinfo {author} {\bibfnamefont {T.}~\bibnamefont {Prosen}},\ and\
  \bibinfo {author} {\bibfnamefont {A.}~\bibnamefont {Chan}},\ }\bibfield
  {title} {\bibinfo {title} {Spectral statistics of non-hermitian matrices and
  dissipative quantum chaos},\ }\href
  {https://doi.org/10.1103/PhysRevLett.127.170602} {\bibfield  {journal}
  {\bibinfo  {journal} {Phys. Rev. Lett.}\ }\textbf {\bibinfo {volume} {127}},\
  \bibinfo {pages} {170602} (\bibinfo {year} {2021})}\BibitemShut {NoStop}%
\bibitem [{\citenamefont {Altland}\ \emph {et~al.}(2021)\citenamefont
  {Altland}, \citenamefont {Fleischhauer},\ and\ \citenamefont
  {Diehl}}]{altland2021symmetry}%
  \BibitemOpen
  \bibfield  {author} {\bibinfo {author} {\bibfnamefont {A.}~\bibnamefont
  {Altland}}, \bibinfo {author} {\bibfnamefont {M.}~\bibnamefont
  {Fleischhauer}},\ and\ \bibinfo {author} {\bibfnamefont {S.}~\bibnamefont
  {Diehl}},\ }\bibfield  {title} {\bibinfo {title} {Symmetry classes of open
  fermionic quantum matter},\ }\href@noop {} {\bibfield  {journal} {\bibinfo
  {journal} {Physical Review X}\ }\textbf {\bibinfo {volume} {11}},\ \bibinfo
  {pages} {021037} (\bibinfo {year} {2021})}\BibitemShut {NoStop}%
\bibitem [{\citenamefont {Costa}\ \emph {et~al.}(2023)\citenamefont {Costa},
  \citenamefont {Ribeiro}, \citenamefont {De~Luca}, \citenamefont {Prosen},\
  and\ \citenamefont {Sá}}]{costa2023}%
  \BibitemOpen
  \bibfield  {author} {\bibinfo {author} {\bibfnamefont {J.}~\bibnamefont
  {Costa}}, \bibinfo {author} {\bibfnamefont {P.}~\bibnamefont {Ribeiro}},
  \bibinfo {author} {\bibfnamefont {A.}~\bibnamefont {De~Luca}}, \bibinfo
  {author} {\bibfnamefont {T.}~\bibnamefont {Prosen}},\ and\ \bibinfo {author}
  {\bibfnamefont {L.}~\bibnamefont {Sá}},\ }\bibfield  {title} {\bibinfo
  {title} {Spectral and steady-state properties of fermionic random quadratic
  liouvillians},\ }\bibfield  {journal} {\bibinfo  {journal} {SciPost Physics}\
  }\textbf {\bibinfo {volume} {15}},\ \href
  {https://doi.org/10.21468/scipostphys.15.4.145}
  {10.21468/scipostphys.15.4.145} (\bibinfo {year} {2023})\BibitemShut
  {NoStop}%
\bibitem [{\citenamefont {Kawabata}\ \emph {et~al.}(2019)\citenamefont
  {Kawabata}, \citenamefont {Shiozaki}, \citenamefont {Ueda},\ and\
  \citenamefont {Sato}}]{ueda2019}%
  \BibitemOpen
  \bibfield  {author} {\bibinfo {author} {\bibfnamefont {K.}~\bibnamefont
  {Kawabata}}, \bibinfo {author} {\bibfnamefont {K.}~\bibnamefont {Shiozaki}},
  \bibinfo {author} {\bibfnamefont {M.}~\bibnamefont {Ueda}},\ and\ \bibinfo
  {author} {\bibfnamefont {M.}~\bibnamefont {Sato}},\ }\bibfield  {title}
  {\bibinfo {title} {Symmetry and topology in non-hermitian physics},\ }\href
  {https://doi.org/10.1103/PhysRevX.9.041015} {\bibfield  {journal} {\bibinfo
  {journal} {Phys. Rev. X}\ }\textbf {\bibinfo {volume} {9}},\ \bibinfo {pages}
  {041015} (\bibinfo {year} {2019})},\ \Eprint
  {https://arxiv.org/abs/1812.09133} {arXiv:1812.09133 [cond-mat.mes-hall]}
  \BibitemShut {NoStop}%
\bibitem [{\citenamefont {Ginibre}(1965)}]{ginibre1965}%
  \BibitemOpen
  \bibfield  {author} {\bibinfo {author} {\bibfnamefont {J.}~\bibnamefont
  {Ginibre}},\ }\bibfield  {title} {\bibinfo {title} {Statistical ensembles of
  complex, quaternion, and real matrices},\ }\href
  {https://doi.org/10.1063/1.1704292} {\bibfield  {journal} {\bibinfo
  {journal} {Journal of Mathematical Physics}\ }\textbf {\bibinfo {volume}
  {6}},\ \bibinfo {pages} {440} (\bibinfo {year} {1965})}\BibitemShut {NoStop}%
\bibitem [{\citenamefont {Garc\'{\i}a-Garc\'{\i}a}\ \emph
  {et~al.}(2002)\citenamefont {Garc\'{\i}a-Garc\'{\i}a}, \citenamefont
  {Nishigaki},\ and\ \citenamefont {Verbaarschot}}]{garcia2002a}%
  \BibitemOpen
  \bibfield  {author} {\bibinfo {author} {\bibfnamefont {A.~M.}\ \bibnamefont
  {Garc\'{\i}a-Garc\'{\i}a}}, \bibinfo {author} {\bibfnamefont {S.~M.}\
  \bibnamefont {Nishigaki}},\ and\ \bibinfo {author} {\bibfnamefont {J.~J.~M.}\
  \bibnamefont {Verbaarschot}},\ }\bibfield  {title} {\bibinfo {title}
  {Critical statistics for non-hermitian matrices},\ }\href
  {https://doi.org/10.1103/PhysRevE.66.016132} {\bibfield  {journal} {\bibinfo
  {journal} {Phys. Rev. E}\ }\textbf {\bibinfo {volume} {66}},\ \bibinfo
  {pages} {016132} (\bibinfo {year} {2002})}\BibitemShut {NoStop}%
\bibitem [{\citenamefont {Garc\'{\i}a-Garc\'{\i}a}\ \emph
  {et~al.}(2023{\natexlab{a}})\citenamefont {Garc\'{\i}a-Garc\'{\i}a},
  \citenamefont {S\'a},\ and\ \citenamefont {Verbaarschot}}]{garcia2023e}%
  \BibitemOpen
  \bibfield  {author} {\bibinfo {author} {\bibfnamefont {A.~M.}\ \bibnamefont
  {Garc\'{\i}a-Garc\'{\i}a}}, \bibinfo {author} {\bibfnamefont
  {L.}~\bibnamefont {S\'a}},\ and\ \bibinfo {author} {\bibfnamefont {J.~J.~M.}\
  \bibnamefont {Verbaarschot}},\ }\bibfield  {title} {\bibinfo {title}
  {Universality and its limits in non-hermitian many-body quantum chaos using
  the {Sachdev-Ye-Kitaev model}},\ }\href
  {https://doi.org/10.1103/PhysRevD.107.066007} {\bibfield  {journal} {\bibinfo
   {journal} {Phys. Rev. D}\ }\textbf {\bibinfo {volume} {107}},\ \bibinfo
  {pages} {066007} (\bibinfo {year} {2023}{\natexlab{a}})}\BibitemShut
  {NoStop}%
\bibitem [{\citenamefont {S\'a}\ \emph {et~al.}(2022)\citenamefont {S\'a},
  \citenamefont {Ribeiro},\ and\ \citenamefont {Prosen}}]{sa2022}%
  \BibitemOpen
  \bibfield  {author} {\bibinfo {author} {\bibfnamefont {L.}~\bibnamefont
  {S\'a}}, \bibinfo {author} {\bibfnamefont {P.}~\bibnamefont {Ribeiro}},\ and\
  \bibinfo {author} {\bibfnamefont {T.}~\bibnamefont {Prosen}},\ }\bibfield
  {title} {\bibinfo {title} {Lindbladian dissipation of strongly-correlated
  quantum matter},\ }\href {https://doi.org/10.1103/PhysRevResearch.4.L022068}
  {\bibfield  {journal} {\bibinfo  {journal} {Phys. Rev. Research}\ }\textbf
  {\bibinfo {volume} {4}},\ \bibinfo {pages} {L022068} (\bibinfo {year}
  {2022})}\BibitemShut {NoStop}%
\bibitem [{\citenamefont {Kulkarni}\ \emph {et~al.}(2022)\citenamefont
  {Kulkarni}, \citenamefont {Numasawa},\ and\ \citenamefont
  {Ryu}}]{kulkarni2022}%
  \BibitemOpen
  \bibfield  {author} {\bibinfo {author} {\bibfnamefont {A.}~\bibnamefont
  {Kulkarni}}, \bibinfo {author} {\bibfnamefont {T.}~\bibnamefont {Numasawa}},\
  and\ \bibinfo {author} {\bibfnamefont {S.}~\bibnamefont {Ryu}},\ }\bibfield
  {title} {\bibinfo {title} {{Lindbladian dynamics of the Sachdev-Ye-Kitaev
  model}},\ }\href {https://doi.org/10.1103/PhysRevB.106.075138} {\bibfield
  {journal} {\bibinfo  {journal} {Phys. Rev. B}\ }\textbf {\bibinfo {volume}
  {106}},\ \bibinfo {pages} {075138} (\bibinfo {year} {2022})}\BibitemShut
  {NoStop}%
\bibitem [{\citenamefont {Garc\'{\i}a-Garc\'{\i}a}\ \emph
  {et~al.}(2023{\natexlab{b}})\citenamefont {Garc\'{\i}a-Garc\'{\i}a},
  \citenamefont {S\'a}, \citenamefont {Verbaarschot},\ and\ \citenamefont
  {Zheng}}]{garcia2022e}%
  \BibitemOpen
  \bibfield  {author} {\bibinfo {author} {\bibfnamefont {A.~M.}\ \bibnamefont
  {Garc\'{\i}a-Garc\'{\i}a}}, \bibinfo {author} {\bibfnamefont
  {L.}~\bibnamefont {S\'a}}, \bibinfo {author} {\bibfnamefont {J.~J.~M.}\
  \bibnamefont {Verbaarschot}},\ and\ \bibinfo {author} {\bibfnamefont {J.~P.}\
  \bibnamefont {Zheng}},\ }\bibfield  {title} {\bibinfo {title} {Keldysh
  wormholes and anomalous relaxation in the dissipative {Sachdev-Ye-Kitaev
  model}},\ }\href {https://doi.org/10.1103/PhysRevD.107.106006} {\bibfield
  {journal} {\bibinfo  {journal} {Phys. Rev. D}\ }\textbf {\bibinfo {volume}
  {107}},\ \bibinfo {pages} {106006} (\bibinfo {year}
  {2023}{\natexlab{b}})}\BibitemShut {NoStop}%
\bibitem [{\citenamefont {Belavin}\ \emph {et~al.}(1969)\citenamefont
  {Belavin}, \citenamefont {Zeldovich}, \citenamefont {Perelomov},\ and\
  \citenamefont {Popov}}]{belavin1969}%
  \BibitemOpen
  \bibfield  {author} {\bibinfo {author} {\bibfnamefont {A.~A.}\ \bibnamefont
  {Belavin}}, \bibinfo {author} {\bibfnamefont {B.~Y.}\ \bibnamefont
  {Zeldovich}}, \bibinfo {author} {\bibfnamefont {A.~M.}\ \bibnamefont
  {Perelomov}},\ and\ \bibinfo {author} {\bibfnamefont {V.~S.}\ \bibnamefont
  {Popov}},\ }\bibfield  {title} {\bibinfo {title} {{Relaxation of quantum
  systems with equidistant spectra}},\ }\href
  {http://jetp.ras.ru/cgi-bin/e/index/e/29/1/p145?a=list} {\bibfield  {journal}
  {\bibinfo  {journal} {Sov. Phys. JETP}\ }\textbf {\bibinfo {volume} {29}},\
  \bibinfo {pages} {145} (\bibinfo {year} {1969})}\BibitemShut {NoStop}%
\bibitem [{\citenamefont {Lindblad}(1976)}]{lindblad1976}%
  \BibitemOpen
  \bibfield  {author} {\bibinfo {author} {\bibfnamefont {G.}~\bibnamefont
  {Lindblad}},\ }\bibfield  {title} {\bibinfo {title} {{On the generators of
  quantum dynamical semigroups}},\ }\href
  {https://doi.org/https://doi.org/10.1007/BF01608499} {\bibfield  {journal}
  {\bibinfo  {journal} {Commun. Math. Phys.}\ }\textbf {\bibinfo {volume}
  {48}},\ \bibinfo {pages} {119} (\bibinfo {year} {1976})}\BibitemShut
  {NoStop}%
\bibitem [{\citenamefont {Gorini}\ \emph {et~al.}(1976)\citenamefont {Gorini},
  \citenamefont {Kossakowski},\ and\ \citenamefont {Sudarshan}}]{gorini1976}%
  \BibitemOpen
  \bibfield  {author} {\bibinfo {author} {\bibfnamefont {V.}~\bibnamefont
  {Gorini}}, \bibinfo {author} {\bibfnamefont {A.}~\bibnamefont
  {Kossakowski}},\ and\ \bibinfo {author} {\bibfnamefont {E.~C.~G.}\
  \bibnamefont {Sudarshan}},\ }\bibfield  {title} {\bibinfo {title}
  {{Completely positive dynamical semigroups of $N$-level systems}},\ }\href
  {https://doi.org/https://doi.org/10.1063/1.522979} {\bibfield  {journal}
  {\bibinfo  {journal} {J. Math. Phys.}\ }\textbf {\bibinfo {volume} {17}},\
  \bibinfo {pages} {821} (\bibinfo {year} {1976})}\BibitemShut {NoStop}%
\bibitem [{\citenamefont {Breuer}\ \emph {et~al.}(2002)\citenamefont {Breuer},
  \citenamefont {Petruccione} \emph {et~al.}}]{breuer2002}%
  \BibitemOpen
  \bibfield  {author} {\bibinfo {author} {\bibfnamefont {H.-P.}\ \bibnamefont
  {Breuer}}, \bibinfo {author} {\bibfnamefont {F.}~\bibnamefont {Petruccione}},
  \emph {et~al.},\ }\href@noop {} {\emph {\bibinfo {title} {The theory of open
  quantum systems}}}\ (\bibinfo  {publisher} {Oxford University Press on
  Demand},\ \bibinfo {year} {2002})\BibitemShut {NoStop}%
\bibitem [{\citenamefont {Manzano}(2020)}]{manzano2020}%
  \BibitemOpen
  \bibfield  {author} {\bibinfo {author} {\bibfnamefont {D.}~\bibnamefont
  {Manzano}},\ }\bibfield  {title} {\bibinfo {title} {{A short introduction to
  the Lindblad master equation}},\ }\href {https://doi.org/10.1063/1.5115323}
  {\bibfield  {journal} {\bibinfo  {journal} {AIP Advances}\ }\textbf {\bibinfo
  {volume} {10}},\ \bibinfo {pages} {025106} (\bibinfo {year}
  {2020})}\BibitemShut {NoStop}%
\bibitem [{\citenamefont {Bergamasco}\ \emph {et~al.}(2023)\citenamefont
  {Bergamasco}, \citenamefont {Carlo},\ and\ \citenamefont
  {Rivas}}]{bergamasco2023quantum}%
  \BibitemOpen
  \bibfield  {author} {\bibinfo {author} {\bibfnamefont {P.~D.}\ \bibnamefont
  {Bergamasco}}, \bibinfo {author} {\bibfnamefont {G.~G.}\ \bibnamefont
  {Carlo}},\ and\ \bibinfo {author} {\bibfnamefont {A.~M.}\ \bibnamefont
  {Rivas}},\ }\bibfield  {title} {\bibinfo {title} {Quantum lyapunov exponent
  in dissipative systems},\ }\href@noop {} {\bibfield  {journal} {\bibinfo
  {journal} {Physical Review E}\ }\textbf {\bibinfo {volume} {108}},\ \bibinfo
  {pages} {024208} (\bibinfo {year} {2023})}\BibitemShut {NoStop}%
\bibitem [{\citenamefont {Yoshida}\ and\ \citenamefont
  {Yao}(2019)}]{yoshida2019disentangling}%
  \BibitemOpen
  \bibfield  {author} {\bibinfo {author} {\bibfnamefont {B.}~\bibnamefont
  {Yoshida}}\ and\ \bibinfo {author} {\bibfnamefont {N.~Y.}\ \bibnamefont
  {Yao}},\ }\bibfield  {title} {\bibinfo {title} {Disentangling scrambling and
  decoherence via quantum teleportation},\ }\href@noop {} {\bibfield  {journal}
  {\bibinfo  {journal} {Physical Review X}\ }\textbf {\bibinfo {volume} {9}},\
  \bibinfo {pages} {011006} (\bibinfo {year} {2019})}\BibitemShut {NoStop}%
\bibitem [{\citenamefont {Tuziemski}(2019)}]{tuziemski2019out}%
  \BibitemOpen
  \bibfield  {author} {\bibinfo {author} {\bibfnamefont {J.}~\bibnamefont
  {Tuziemski}},\ }\bibfield  {title} {\bibinfo {title} {Out-of-time-ordered
  correlation functions in open systems: A {Feynman-Vernon} influence
  functional approach},\ }\href@noop {} {\bibfield  {journal} {\bibinfo
  {journal} {Physical Review A}\ }\textbf {\bibinfo {volume} {100}},\ \bibinfo
  {pages} {062106} (\bibinfo {year} {2019})}\BibitemShut {NoStop}%
\bibitem [{\citenamefont {Weinstein}\ \emph {et~al.}(2023)\citenamefont
  {Weinstein}, \citenamefont {Kelly}, \citenamefont {Marino},\ and\
  \citenamefont {Altman}}]{weinstein2023scrambling}%
  \BibitemOpen
  \bibfield  {author} {\bibinfo {author} {\bibfnamefont {Z.}~\bibnamefont
  {Weinstein}}, \bibinfo {author} {\bibfnamefont {S.~P.}\ \bibnamefont
  {Kelly}}, \bibinfo {author} {\bibfnamefont {J.}~\bibnamefont {Marino}},\ and\
  \bibinfo {author} {\bibfnamefont {E.}~\bibnamefont {Altman}},\ }\bibfield
  {title} {\bibinfo {title} {Scrambling transition in a radiative random
  unitary circuit},\ }\href@noop {} {\bibfield  {journal} {\bibinfo  {journal}
  {Physical Review Letters}\ }\textbf {\bibinfo {volume} {131}},\ \bibinfo
  {pages} {220404} (\bibinfo {year} {2023})}\BibitemShut {NoStop}%
\bibitem [{\citenamefont {Zanardi}\ and\ \citenamefont
  {Anand}(2021)}]{zanardi2021information}%
  \BibitemOpen
  \bibfield  {author} {\bibinfo {author} {\bibfnamefont {P.}~\bibnamefont
  {Zanardi}}\ and\ \bibinfo {author} {\bibfnamefont {N.}~\bibnamefont
  {Anand}},\ }\bibfield  {title} {\bibinfo {title} {Information scrambling and
  chaos in open quantum systems},\ }\href@noop {} {\bibfield  {journal}
  {\bibinfo  {journal} {Physical Review A}\ }\textbf {\bibinfo {volume}
  {103}},\ \bibinfo {pages} {062214} (\bibinfo {year} {2021})}\BibitemShut
  {NoStop}%
\bibitem [{\citenamefont {qian Huang}\ \emph {et~al.}(2020)\citenamefont {qian
  Huang}, \citenamefont {Wang}, \citenamefont {Zhao},\ and\ \citenamefont
  {Liu}}]{huang2020}%
  \BibitemOpen
  \bibfield  {author} {\bibinfo {author} {\bibfnamefont {K.}~\bibnamefont {qian
  Huang}}, \bibinfo {author} {\bibfnamefont {J.}~\bibnamefont {Wang}}, \bibinfo
  {author} {\bibfnamefont {W.-L.}\ \bibnamefont {Zhao}},\ and\ \bibinfo
  {author} {\bibfnamefont {J.}~\bibnamefont {Liu}},\ }\bibfield  {title}
  {\bibinfo {title} {Chaotic dynamics of a non-hermitian kicked particle},\
  }\href {https://doi.org/10.1088/1361-648X/abbcf8} {\bibfield  {journal}
  {\bibinfo  {journal} {Journal of Physics: Condensed Matter}\ }\textbf
  {\bibinfo {volume} {33}},\ \bibinfo {pages} {055402} (\bibinfo {year}
  {2020})}\BibitemShut {NoStop}%
\bibitem [{\citenamefont {Syzranov}\ \emph {et~al.}(2018)\citenamefont
  {Syzranov}, \citenamefont {Gorshkov},\ and\ \citenamefont
  {Galitski}}]{syzranov2018out}%
  \BibitemOpen
  \bibfield  {author} {\bibinfo {author} {\bibfnamefont {S.}~\bibnamefont
  {Syzranov}}, \bibinfo {author} {\bibfnamefont {A.}~\bibnamefont {Gorshkov}},\
  and\ \bibinfo {author} {\bibfnamefont {V.}~\bibnamefont {Galitski}},\
  }\bibfield  {title} {\bibinfo {title} {Out-of-time-order correlators in
  finite open systems},\ }\href@noop {} {\bibfield  {journal} {\bibinfo
  {journal} {Physical Review B}\ }\textbf {\bibinfo {volume} {97}},\ \bibinfo
  {pages} {161114} (\bibinfo {year} {2018})}\BibitemShut {NoStop}%
\bibitem [{\citenamefont {Zhai}\ and\ \citenamefont {Yin}(2020)}]{zhai2020a}%
  \BibitemOpen
  \bibfield  {author} {\bibinfo {author} {\bibfnamefont {L.-J.}\ \bibnamefont
  {Zhai}}\ and\ \bibinfo {author} {\bibfnamefont {S.}~\bibnamefont {Yin}},\
  }\bibfield  {title} {\bibinfo {title} {Out-of-time-ordered correlator in
  non-hermitian quantum systems},\ }\href
  {https://doi.org/10.1103/PhysRevB.102.054303} {\bibfield  {journal} {\bibinfo
   {journal} {Phys. Rev. B}\ }\textbf {\bibinfo {volume} {102}},\ \bibinfo
  {pages} {054303} (\bibinfo {year} {2020})}\BibitemShut {NoStop}%
\bibitem [{\citenamefont {Turiaci}(2019)}]{turiaci2019a}%
  \BibitemOpen
  \bibfield  {author} {\bibinfo {author} {\bibfnamefont {G.~J.}\ \bibnamefont
  {Turiaci}},\ }\bibfield  {title} {\bibinfo {title} {An inelastic bound on
  chaos},\ }\bibfield  {journal} {\bibinfo  {journal} {Journal of High Energy
  Physics}\ }\textbf {\bibinfo {volume} {2019}},\ \href
  {https://doi.org/10.1007/jhep07(2019)099} {10.1007/jhep07(2019)099} (\bibinfo
  {year} {2019})\BibitemShut {NoStop}%
\bibitem [{\citenamefont {Garc\'{\i}a-Garc\'{\i}a}\ \emph
  {et~al.}(2022{\natexlab{b}})\citenamefont {Garc\'{\i}a-Garc\'{\i}a},
  \citenamefont {Godet}, \citenamefont {Yin},\ and\ \citenamefont
  {Zheng}}]{garcia2022c}%
  \BibitemOpen
  \bibfield  {author} {\bibinfo {author} {\bibfnamefont {A.~M.}\ \bibnamefont
  {Garc\'{\i}a-Garc\'{\i}a}}, \bibinfo {author} {\bibfnamefont
  {V.}~\bibnamefont {Godet}}, \bibinfo {author} {\bibfnamefont
  {C.}~\bibnamefont {Yin}},\ and\ \bibinfo {author} {\bibfnamefont {J.~P.}\
  \bibnamefont {Zheng}},\ }\bibfield  {title} {\bibinfo {title}
  {{Euclidean-to-Lorentzian wormhole transition and gravitational symmetry
  breaking in the Sachdev-Ye-Kitaev model}},\ }\href
  {https://doi.org/10.1103/PhysRevD.106.046008} {\bibfield  {journal} {\bibinfo
   {journal} {Phys. Rev. D}\ }\textbf {\bibinfo {volume} {106}},\ \bibinfo
  {pages} {046008} (\bibinfo {year} {2022}{\natexlab{b}})}\BibitemShut
  {NoStop}%
\bibitem [{\citenamefont {Cai}\ \emph {et~al.}(2022)\citenamefont {Cai},
  \citenamefont {Cao}, \citenamefont {Ge}, \citenamefont {Matsumoto},\ and\
  \citenamefont {Sin}}]{Cai2022}%
  \BibitemOpen
  \bibfield  {author} {\bibinfo {author} {\bibfnamefont {W.}~\bibnamefont
  {Cai}}, \bibinfo {author} {\bibfnamefont {S.}~\bibnamefont {Cao}}, \bibinfo
  {author} {\bibfnamefont {X.-H.}\ \bibnamefont {Ge}}, \bibinfo {author}
  {\bibfnamefont {M.}~\bibnamefont {Matsumoto}},\ and\ \bibinfo {author}
  {\bibfnamefont {S.-J.}\ \bibnamefont {Sin}},\ }\bibfield  {title} {\bibinfo
  {title} {Non-hermitian quantum system generated from two coupled
  sachdev-ye-kitaev models},\ }\href
  {https://doi.org/10.1103/PhysRevD.106.106010} {\bibfield  {journal} {\bibinfo
   {journal} {Phys. Rev. D}\ }\textbf {\bibinfo {volume} {106}},\ \bibinfo
  {pages} {106010} (\bibinfo {year} {2022})}\BibitemShut {NoStop}%
\bibitem [{\citenamefont {Schuster}\ and\ \citenamefont
  {Yao}(2023)}]{schuster:2022bot}%
  \BibitemOpen
  \bibfield  {author} {\bibinfo {author} {\bibfnamefont {T.}~\bibnamefont
  {Schuster}}\ and\ \bibinfo {author} {\bibfnamefont {N.~Y.}\ \bibnamefont
  {Yao}},\ }\bibfield  {title} {\bibinfo {title} {{Operator Growth in Open
  Quantum Systems}},\ }\href {https://doi.org/10.1103/PhysRevLett.131.160402}
  {\bibfield  {journal} {\bibinfo  {journal} {Phys. Rev. Lett.}\ }\textbf
  {\bibinfo {volume} {131}},\ \bibinfo {pages} {160402} (\bibinfo {year}
  {2023})},\ \Eprint {https://arxiv.org/abs/2208.12272} {arXiv:2208.12272
  [quant-ph]} \BibitemShut {NoStop}%
\bibitem [{\citenamefont {Bhattacharya}\ \emph {et~al.}(2022)\citenamefont
  {Bhattacharya}, \citenamefont {Nandy}, \citenamefont {Nath},\ and\
  \citenamefont {Sahu}}]{bhattacharya2022}%
  \BibitemOpen
  \bibfield  {author} {\bibinfo {author} {\bibfnamefont {A.}~\bibnamefont
  {Bhattacharya}}, \bibinfo {author} {\bibfnamefont {P.}~\bibnamefont {Nandy}},
  \bibinfo {author} {\bibfnamefont {P.~P.}\ \bibnamefont {Nath}},\ and\
  \bibinfo {author} {\bibfnamefont {H.}~\bibnamefont {Sahu}},\ }\bibfield
  {title} {\bibinfo {title} {{Operator growth and Krylov construction in
  dissipative open quantum systems}},\ }\bibfield  {journal} {\bibinfo
  {journal} {Journal of High Energy Physics}\ }\textbf {\bibinfo {volume}
  {2022}},\ \href {https://doi.org/10.1007/jhep12(2022)081}
  {10.1007/jhep12(2022)081} (\bibinfo {year} {2022})\BibitemShut {NoStop}%
\bibitem [{\citenamefont {Bhattacharjee}\ \emph
  {et~al.}(2023{\natexlab{a}})\citenamefont {Bhattacharjee}, \citenamefont
  {Cao}, \citenamefont {Nandy},\ and\ \citenamefont
  {Pathak}}]{bhattacharjee2023}%
  \BibitemOpen
  \bibfield  {author} {\bibinfo {author} {\bibfnamefont {B.}~\bibnamefont
  {Bhattacharjee}}, \bibinfo {author} {\bibfnamefont {X.}~\bibnamefont {Cao}},
  \bibinfo {author} {\bibfnamefont {P.}~\bibnamefont {Nandy}},\ and\ \bibinfo
  {author} {\bibfnamefont {T.}~\bibnamefont {Pathak}},\ }\bibfield  {title}
  {\bibinfo {title} {Operator growth in open quantum systems: lessons from the
  dissipative {SYK}},\ }\bibfield  {journal} {\bibinfo  {journal} {Journal of
  High Energy Physics}\ }\textbf {\bibinfo {volume} {2023}},\ \href
  {https://doi.org/10.1007/jhep03(2023)054} {10.1007/jhep03(2023)054} (\bibinfo
  {year} {2023}{\natexlab{a}})\BibitemShut {NoStop}%
\bibitem [{\citenamefont {Zhang}\ \emph {et~al.}(2021)\citenamefont {Zhang},
  \citenamefont {Jian}, \citenamefont {Liu},\ and\ \citenamefont
  {Chen}}]{pengfei2021}%
  \BibitemOpen
  \bibfield  {author} {\bibinfo {author} {\bibfnamefont {P.}~\bibnamefont
  {Zhang}}, \bibinfo {author} {\bibfnamefont {S.-K.}\ \bibnamefont {Jian}},
  \bibinfo {author} {\bibfnamefont {C.}~\bibnamefont {Liu}},\ and\ \bibinfo
  {author} {\bibfnamefont {X.}~\bibnamefont {Chen}},\ }\bibfield  {title}
  {\bibinfo {title} {Emergent {R}eplica {C}onformal {S}ymmetry in
  {N}on-{H}ermitian {SYK}{$_2$} {C}hains},\ }\href
  {https://doi.org/10.22331/q-2021-11-16-579} {\bibfield  {journal} {\bibinfo
  {journal} {{Quantum}}\ }\textbf {\bibinfo {volume} {5}},\ \bibinfo {pages}
  {579} (\bibinfo {year} {2021})}\BibitemShut {NoStop}%
\bibitem [{\citenamefont {Jian}\ \emph {et~al.}(2021)\citenamefont {Jian},
  \citenamefont {Liu}, \citenamefont {Chen}, \citenamefont {Swingle},\ and\
  \citenamefont {Zhang}}]{pengfei2021c}%
  \BibitemOpen
  \bibfield  {author} {\bibinfo {author} {\bibfnamefont {S.-K.}\ \bibnamefont
  {Jian}}, \bibinfo {author} {\bibfnamefont {C.}~\bibnamefont {Liu}}, \bibinfo
  {author} {\bibfnamefont {X.}~\bibnamefont {Chen}}, \bibinfo {author}
  {\bibfnamefont {B.}~\bibnamefont {Swingle}},\ and\ \bibinfo {author}
  {\bibfnamefont {P.}~\bibnamefont {Zhang}},\ }\bibfield  {title} {\bibinfo
  {title} {{Measurement-Induced Phase Transition in the Monitored
  Sachdev-Ye-Kitaev Model}},\ }\href
  {https://doi.org/10.1103/PhysRevLett.127.140601} {\bibfield  {journal}
  {\bibinfo  {journal} {Phys. Rev. Lett.}\ }\textbf {\bibinfo {volume} {127}},\
  \bibinfo {pages} {140601} (\bibinfo {year} {2021})},\ \Eprint
  {https://arxiv.org/abs/2106.09635} {arXiv:2106.09635 [quant-ph]} \BibitemShut
  {NoStop}%
\bibitem [{\citenamefont {Bhattacharjee}\ \emph
  {et~al.}(2023{\natexlab{b}})\citenamefont {Bhattacharjee}, \citenamefont
  {Nandy},\ and\ \citenamefont {Pathak}}]{bhattacharjee2023a}%
  \BibitemOpen
  \bibfield  {author} {\bibinfo {author} {\bibfnamefont {B.}~\bibnamefont
  {Bhattacharjee}}, \bibinfo {author} {\bibfnamefont {P.}~\bibnamefont
  {Nandy}},\ and\ \bibinfo {author} {\bibfnamefont {T.}~\bibnamefont
  {Pathak}},\ }\href@noop {} {\bibinfo {title} {Operator dynamics in
  {Lindbladian SYK: a Krylov complexity} perspective}} (\bibinfo {year}
  {2023}{\natexlab{b}}),\ \Eprint {https://arxiv.org/abs/2311.00753}
  {arXiv:2311.00753 [quant-ph]} \BibitemShut {NoStop}%
\bibitem [{\citenamefont {Liu}\ \emph {et~al.}(2023)\citenamefont {Liu},
  \citenamefont {Tang},\ and\ \citenamefont {Zhai}}]{liu2023}%
  \BibitemOpen
  \bibfield  {author} {\bibinfo {author} {\bibfnamefont {C.}~\bibnamefont
  {Liu}}, \bibinfo {author} {\bibfnamefont {H.}~\bibnamefont {Tang}},\ and\
  \bibinfo {author} {\bibfnamefont {H.}~\bibnamefont {Zhai}},\ }\bibfield
  {title} {\bibinfo {title} {{Krylov complexity in open quantum systems}},\
  }\href {https://doi.org/10.1103/PhysRevResearch.5.033085} {\bibfield
  {journal} {\bibinfo  {journal} {Phys. Rev. Res.}\ }\textbf {\bibinfo {volume}
  {5}},\ \bibinfo {pages} {033085} (\bibinfo {year} {2023})}\BibitemShut
  {NoStop}%
\bibitem [{\citenamefont {Srivatsa}\ and\ \citenamefont {von
  Keyserlingk}(2023)}]{srivatsa2023}%
  \BibitemOpen
  \bibfield  {author} {\bibinfo {author} {\bibfnamefont {N.~S.}\ \bibnamefont
  {Srivatsa}}\ and\ \bibinfo {author} {\bibfnamefont {C.}~\bibnamefont {von
  Keyserlingk}},\ }\href@noop {} {\bibinfo {title} {The operator growth
  hypothesis in open quantum systems}} (\bibinfo {year} {2023}),\ \Eprint
  {https://arxiv.org/abs/2310.15376} {arXiv:2310.15376 [quant-ph]} \BibitemShut
  {NoStop}%
\bibitem [{\citenamefont {Zhang}\ and\ \citenamefont {Yu}(2023)}]{pengfei2023}%
  \BibitemOpen
  \bibfield  {author} {\bibinfo {author} {\bibfnamefont {P.}~\bibnamefont
  {Zhang}}\ and\ \bibinfo {author} {\bibfnamefont {Z.}~\bibnamefont {Yu}},\
  }\bibfield  {title} {\bibinfo {title} {Dynamical transition of operator size
  growth in quantum systems embedded in an environment},\ }\href
  {https://doi.org/10.1103/PhysRevLett.130.250401} {\bibfield  {journal}
  {\bibinfo  {journal} {Phys. Rev. Lett.}\ }\textbf {\bibinfo {volume} {130}},\
  \bibinfo {pages} {250401} (\bibinfo {year} {2023})}\BibitemShut {NoStop}%
\bibitem [{\citenamefont {Liu}\ \emph {et~al.}(2024)\citenamefont {Liu},
  \citenamefont {Meyer},\ and\ \citenamefont {Xian}}]{liu2024}%
  \BibitemOpen
  \bibfield  {author} {\bibinfo {author} {\bibfnamefont {J.}~\bibnamefont
  {Liu}}, \bibinfo {author} {\bibfnamefont {R.}~\bibnamefont {Meyer}},\ and\
  \bibinfo {author} {\bibfnamefont {Z.-Y.}\ \bibnamefont {Xian}},\ }\href@noop
  {} {\bibinfo {title} {{Operator size growth in Lindbladian SYK}}} (\bibinfo
  {year} {2024}),\ \Eprint {https://arxiv.org/abs/2403.07115} {arXiv:2403.07115
  [hep-th]} \BibitemShut {NoStop}%
\bibitem [{\citenamefont {Chen}\ \emph {et~al.}(2017)\citenamefont {Chen},
  \citenamefont {Zhai},\ and\ \citenamefont {Zhang}}]{chen2017}%
  \BibitemOpen
  \bibfield  {author} {\bibinfo {author} {\bibfnamefont {Y.}~\bibnamefont
  {Chen}}, \bibinfo {author} {\bibfnamefont {H.}~\bibnamefont {Zhai}},\ and\
  \bibinfo {author} {\bibfnamefont {P.}~\bibnamefont {Zhang}},\ }\bibfield
  {title} {\bibinfo {title} {Tunable quantum chaos in the {Sachdev-Ye-Kitaev}
  model coupled to a thermal bath},\ }\bibfield  {journal} {\bibinfo  {journal}
  {Journal of High Energy Physics}\ }\textbf {\bibinfo {volume} {2017}},\ \href
  {https://doi.org/10.1007/jhep07(2017)150} {10.1007/jhep07(2017)150} (\bibinfo
  {year} {2017})\BibitemShut {NoStop}%
\bibitem [{\citenamefont {Maldacena}\ and\ \citenamefont
  {Qi}(2018)}]{maldacena2018}%
  \BibitemOpen
  \bibfield  {author} {\bibinfo {author} {\bibfnamefont {J.}~\bibnamefont
  {Maldacena}}\ and\ \bibinfo {author} {\bibfnamefont {X.-L.}\ \bibnamefont
  {Qi}},\ }\bibfield  {title} {\bibinfo {title} {{Eternal traversable
  wormhole}},\ }\href@noop {} {\bibfield  {journal} {\bibinfo  {journal}
  {eprint}\ } (\bibinfo {year} {2018})},\ \Eprint
  {https://arxiv.org/abs/1804.00491} {arXiv:1804.00491 [hep-th]} \BibitemShut
  {NoStop}%
\bibitem [{\citenamefont {C{\'{a}}ceres}\ \emph {et~al.}(2021)\citenamefont
  {C{\'{a}}ceres}, \citenamefont {Misobuchi},\ and\ \citenamefont
  {Pimentel}}]{caceres2021}%
  \BibitemOpen
  \bibfield  {author} {\bibinfo {author} {\bibfnamefont {E.}~\bibnamefont
  {C{\'{a}}ceres}}, \bibinfo {author} {\bibfnamefont {A.}~\bibnamefont
  {Misobuchi}},\ and\ \bibinfo {author} {\bibfnamefont {R.}~\bibnamefont
  {Pimentel}},\ }\bibfield  {title} {\bibinfo {title} {Sparse {SYK} and
  traversable wormholes},\ }\href {https://doi.org/10.1007/JHEP11(2021)015}
  {\bibfield  {journal} {\bibinfo  {journal} {Journal of High Energy Physics}\
  }\textbf {\bibinfo {volume} {2021}},\ \bibinfo {pages} {11} (\bibinfo {year}
  {2021})},\ \Eprint {https://arxiv.org/abs/2108.08808} {arXiv:2108.08808
  [hep-th]} \BibitemShut {NoStop}%
\bibitem [{\citenamefont {Nosaka}\ and\ \citenamefont
  {Numasawa}(2021)}]{nosaka2021}%
  \BibitemOpen
  \bibfield  {author} {\bibinfo {author} {\bibfnamefont {T.}~\bibnamefont
  {Nosaka}}\ and\ \bibinfo {author} {\bibfnamefont {T.}~\bibnamefont
  {Numasawa}},\ }\bibfield  {title} {\bibinfo {title} {Chaos exponents of {SYK}
  traversable wormholes},\ }\bibfield  {journal} {\bibinfo  {journal} {Journal
  of High Energy Physics}\ }\textbf {\bibinfo {volume} {2021}},\ \href
  {https://doi.org/10.1007/jhep02(2021)150} {10.1007/jhep02(2021)150} (\bibinfo
  {year} {2021})\BibitemShut {NoStop}%
\bibitem [{\citenamefont {Nosaka}\ and\ \citenamefont
  {Numasawa}(2023)}]{nosaka2022}%
  \BibitemOpen
  \bibfield  {author} {\bibinfo {author} {\bibfnamefont {T.}~\bibnamefont
  {Nosaka}}\ and\ \bibinfo {author} {\bibfnamefont {T.}~\bibnamefont
  {Numasawa}},\ }\bibfield  {title} {\bibinfo {title} {On {SYK} traversable
  wormhole with imperfectly correlated disorders},\ }\bibfield  {journal}
  {\bibinfo  {journal} {Journal of High Energy Physics}\ }\textbf {\bibinfo
  {volume} {2023}},\ \href {https://doi.org/10.1007/jhep04(2023)145}
  {10.1007/jhep04(2023)145} (\bibinfo {year} {2023})\BibitemShut {NoStop}%
\bibitem [{\citenamefont {Caldeira}\ and\ \citenamefont
  {Leggett}(1981)}]{caldeira1981}%
  \BibitemOpen
  \bibfield  {author} {\bibinfo {author} {\bibfnamefont {A.~O.}\ \bibnamefont
  {Caldeira}}\ and\ \bibinfo {author} {\bibfnamefont {A.~J.}\ \bibnamefont
  {Leggett}},\ }\bibfield  {title} {\bibinfo {title} {{Influence of Dissipation
  on Quantum Tunneling in Macroscopic Systems}},\ }\href
  {https://doi.org/10.1103/PhysRevLett.46.211} {\bibfield  {journal} {\bibinfo
  {journal} {Phys. Rev. Lett.}\ }\textbf {\bibinfo {volume} {46}},\ \bibinfo
  {pages} {211} (\bibinfo {year} {1981})}\BibitemShut {NoStop}%
\bibitem [{\citenamefont {Kamenev}(2011)}]{kamenevbook}%
  \BibitemOpen
  \bibfield  {author} {\bibinfo {author} {\bibfnamefont {A.}~\bibnamefont
  {Kamenev}},\ }\href@noop {} {\emph {\bibinfo {title} {Field theory of
  non-equilibrium systems}}}\ (\bibinfo  {publisher} {Cambridge University
  Press},\ \bibinfo {year} {2011})\BibitemShut {NoStop}%
\bibitem [{\citenamefont {Sieberer}\ \emph {et~al.}(2016)\citenamefont
  {Sieberer}, \citenamefont {Buchhold},\ and\ \citenamefont
  {Diehl}}]{sieberer2016}%
  \BibitemOpen
  \bibfield  {author} {\bibinfo {author} {\bibfnamefont {L.~M.}\ \bibnamefont
  {Sieberer}}, \bibinfo {author} {\bibfnamefont {M.}~\bibnamefont {Buchhold}},\
  and\ \bibinfo {author} {\bibfnamefont {S.}~\bibnamefont {Diehl}},\ }\bibfield
   {title} {\bibinfo {title} {Keldysh field theory for driven open quantum
  systems},\ }\href {https://doi.org/10.1088/0034-4885/79/9/096001} {\bibfield
  {journal} {\bibinfo  {journal} {Reports on Progress in Physics}\ }\textbf
  {\bibinfo {volume} {79}},\ \bibinfo {pages} {096001} (\bibinfo {year}
  {2016})}\BibitemShut {NoStop}%
\bibitem [{\citenamefont {Garc\'{\i}a-Garc\'{\i}a}\ \emph
  {et~al.}(2022{\natexlab{c}})\citenamefont {Garc\'{\i}a-Garc\'{\i}a},
  \citenamefont {Jia}, \citenamefont {Rosa},\ and\ \citenamefont
  {Verbaarschot}}]{garcia2022b}%
  \BibitemOpen
  \bibfield  {author} {\bibinfo {author} {\bibfnamefont {A.~M.}\ \bibnamefont
  {Garc\'{\i}a-Garc\'{\i}a}}, \bibinfo {author} {\bibfnamefont
  {Y.}~\bibnamefont {Jia}}, \bibinfo {author} {\bibfnamefont {D.}~\bibnamefont
  {Rosa}},\ and\ \bibinfo {author} {\bibfnamefont {J.~J.~M.}\ \bibnamefont
  {Verbaarschot}},\ }\href {https://doi.org/10.48550/ARXIV.2203.13080}
  {\bibinfo {title} {{Replica Symmetry Breaking in Random Non-Hermitian
  Systems}}} (\bibinfo {year} {2022}{\natexlab{c}})\BibitemShut {NoStop}%
\bibitem [{\citenamefont {Kawabata}\ \emph {et~al.}(2023)\citenamefont
  {Kawabata}, \citenamefont {Kulkarni}, \citenamefont {Li}, \citenamefont
  {Numasawa},\ and\ \citenamefont {Ryu}}]{kawabata2022}%
  \BibitemOpen
  \bibfield  {author} {\bibinfo {author} {\bibfnamefont {K.}~\bibnamefont
  {Kawabata}}, \bibinfo {author} {\bibfnamefont {A.}~\bibnamefont {Kulkarni}},
  \bibinfo {author} {\bibfnamefont {J.}~\bibnamefont {Li}}, \bibinfo {author}
  {\bibfnamefont {T.}~\bibnamefont {Numasawa}},\ and\ \bibinfo {author}
  {\bibfnamefont {S.}~\bibnamefont {Ryu}},\ }\bibfield  {title} {\bibinfo
  {title} {{{Dynamical quantum phase transitions in Sachdev-Ye-Kitaev
  Lindbladians}}},\ }\href {https://doi.org/10.1103/PhysRevB.108.075110}
  {\bibfield  {journal} {\bibinfo  {journal} {Phys. Rev. B}\ }\textbf {\bibinfo
  {volume} {108}},\ \bibinfo {pages} {075110} (\bibinfo {year} {2023})},\
  \Eprint {https://arxiv.org/abs/2210.04093} {arXiv:2210.04093
  [cond-mat.stat-mech]} \BibitemShut {NoStop}%
\bibitem [{\citenamefont {Khramtsov}\ and\ \citenamefont
  {Lanina}(2021)}]{khramtsov2021}%
  \BibitemOpen
  \bibfield  {author} {\bibinfo {author} {\bibfnamefont {M.}~\bibnamefont
  {Khramtsov}}\ and\ \bibinfo {author} {\bibfnamefont {E.}~\bibnamefont
  {Lanina}},\ }\bibfield  {title} {\bibinfo {title} {{Spectral form factor in
  the double-scaled SYK model}},\ }\bibfield  {journal} {\bibinfo  {journal}
  {Journal of High Energy Physics}\ }\textbf {\bibinfo {volume} {2021}},\ \href
  {https://doi.org/10.1007/jhep03(2021)031} {10.1007/jhep03(2021)031} (\bibinfo
  {year} {2021})\BibitemShut {NoStop}%
\bibitem [{\citenamefont {Tarnopolsky}(2019)}]{Tarnopolsky:2018env}%
  \BibitemOpen
  \bibfield  {author} {\bibinfo {author} {\bibfnamefont {G.}~\bibnamefont
  {Tarnopolsky}},\ }\bibfield  {title} {\bibinfo {title} {{Large $q$ expansion
  in the Sachdev-Ye-Kitaev model}},\ }\href
  {https://doi.org/10.1103/PhysRevD.99.026010} {\bibfield  {journal} {\bibinfo
  {journal} {Phys. Rev. D}\ }\textbf {\bibinfo {volume} {99}},\ \bibinfo
  {pages} {026010} (\bibinfo {year} {2019})},\ \Eprint
  {https://arxiv.org/abs/1801.06871} {arXiv:1801.06871 [hep-th]} \BibitemShut
  {NoStop}%
\bibitem [{\citenamefont {Lin}\ and\ \citenamefont
  {Motrunich}(2018)}]{lin2018}%
  \BibitemOpen
  \bibfield  {author} {\bibinfo {author} {\bibfnamefont {C.-J.}\ \bibnamefont
  {Lin}}\ and\ \bibinfo {author} {\bibfnamefont {O.~I.}\ \bibnamefont
  {Motrunich}},\ }\bibfield  {title} {\bibinfo {title} {Out-of-time-ordered
  correlators in a quantum {Ising} chain},\ }\href
  {https://doi.org/10.1103/PhysRevB.97.144304} {\bibfield  {journal} {\bibinfo
  {journal} {Phys. Rev. B}\ }\textbf {\bibinfo {volume} {97}},\ \bibinfo
  {pages} {144304} (\bibinfo {year} {2018})}\BibitemShut {NoStop}%
\bibitem [{\citenamefont {Craps}\ \emph {et~al.}(2020)\citenamefont {Craps},
  \citenamefont {De~Clerck}, \citenamefont {Janssens}, \citenamefont {Luyten},\
  and\ \citenamefont {Rabideau}}]{craps2020}%
  \BibitemOpen
  \bibfield  {author} {\bibinfo {author} {\bibfnamefont {B.}~\bibnamefont
  {Craps}}, \bibinfo {author} {\bibfnamefont {M.}~\bibnamefont {De~Clerck}},
  \bibinfo {author} {\bibfnamefont {D.}~\bibnamefont {Janssens}}, \bibinfo
  {author} {\bibfnamefont {V.}~\bibnamefont {Luyten}},\ and\ \bibinfo {author}
  {\bibfnamefont {C.}~\bibnamefont {Rabideau}},\ }\bibfield  {title} {\bibinfo
  {title} {Lyapunov growth in quantum spin chains},\ }\href
  {https://doi.org/10.1103/PhysRevB.101.174313} {\bibfield  {journal} {\bibinfo
   {journal} {Phys. Rev. B}\ }\textbf {\bibinfo {volume} {101}},\ \bibinfo
  {pages} {174313} (\bibinfo {year} {2020})}\BibitemShut {NoStop}%
\bibitem [{\citenamefont {Jalabert}\ \emph {et~al.}(2018)\citenamefont
  {Jalabert}, \citenamefont {Garc\'{\i}a-Mata},\ and\ \citenamefont
  {Wisniacki}}]{jalabert2018}%
  \BibitemOpen
  \bibfield  {author} {\bibinfo {author} {\bibfnamefont {R.~A.}\ \bibnamefont
  {Jalabert}}, \bibinfo {author} {\bibfnamefont {I.}~\bibnamefont
  {Garc\'{\i}a-Mata}},\ and\ \bibinfo {author} {\bibfnamefont {D.~A.}\
  \bibnamefont {Wisniacki}},\ }\bibfield  {title} {\bibinfo {title}
  {Semiclassical theory of out-of-time-order correlators for low-dimensional
  classically chaotic systems},\ }\href
  {https://doi.org/10.1103/PhysRevE.98.062218} {\bibfield  {journal} {\bibinfo
  {journal} {Phys. Rev. E}\ }\textbf {\bibinfo {volume} {98}},\ \bibinfo
  {pages} {062218} (\bibinfo {year} {2018})}\BibitemShut {NoStop}%
\bibitem [{\citenamefont {Garci­a-Mata}\ \emph {et~al.}(2023)\citenamefont
  {Garci­a-Mata}, \citenamefont {Jalabert},\ and\ \citenamefont
  {Wisniacki}}]{garciamata2023}%
  \BibitemOpen
  \bibfield  {author} {\bibinfo {author} {\bibfnamefont {I.}~\bibnamefont
  {Garci­a-Mata}}, \bibinfo {author} {\bibfnamefont {R.}~\bibnamefont
  {Jalabert}},\ and\ \bibinfo {author} {\bibfnamefont {D.}~\bibnamefont
  {Wisniacki}},\ }\bibfield  {title} {\bibinfo {title} {Out-of-time-order
  correlations and quantum chaos},\ }\href
  {https://doi.org/10.4249/scholarpedia.55237} {\bibfield  {journal} {\bibinfo
  {journal} {Scholarpedia}\ }\textbf {\bibinfo {volume} {18}},\ \bibinfo
  {pages} {55237} (\bibinfo {year} {2023})}\BibitemShut {NoStop}%
\bibitem [{\citenamefont {Luitz}\ \emph {et~al.}(2015)\citenamefont {Luitz},
  \citenamefont {Laflorencie},\ and\ \citenamefont {Alet}}]{luitz2015}%
  \BibitemOpen
  \bibfield  {author} {\bibinfo {author} {\bibfnamefont {D.~J.}\ \bibnamefont
  {Luitz}}, \bibinfo {author} {\bibfnamefont {N.}~\bibnamefont {Laflorencie}},\
  and\ \bibinfo {author} {\bibfnamefont {F.}~\bibnamefont {Alet}},\ }\bibfield
  {title} {\bibinfo {title} {Many-body localization edge in the random-field
  {Heisenberg} chain},\ }\href {https://doi.org/10.1103/PhysRevB.91.081103}
  {\bibfield  {journal} {\bibinfo  {journal} {Phys. Rev. B}\ }\textbf {\bibinfo
  {volume} {91}},\ \bibinfo {pages} {081103} (\bibinfo {year} {2015})},\
  \Eprint {https://arxiv.org/abs/1411.0660} {arXiv:1411.0660 [cond-mat.dis-nn]}
  \BibitemShut {NoStop}%
\bibitem [{\citenamefont {Maldacena}\ \emph {et~al.}(2020)\citenamefont
  {Maldacena}, \citenamefont {Turiaci},\ and\ \citenamefont
  {Yang}}]{turiaci2019}%
  \BibitemOpen
  \bibfield  {author} {\bibinfo {author} {\bibfnamefont {J.}~\bibnamefont
  {Maldacena}}, \bibinfo {author} {\bibfnamefont {G.~J.}\ \bibnamefont
  {Turiaci}},\ and\ \bibinfo {author} {\bibfnamefont {Z.}~\bibnamefont
  {Yang}},\ }\href@noop {} {\bibinfo {title} {Two dimensional nearly {de
  Sitter} gravity}} (\bibinfo {year} {2020}),\ \Eprint
  {https://arxiv.org/abs/1904.01911} {arXiv:1904.01911 [hep-th]} \BibitemShut
  {NoStop}%
\bibitem [{\citenamefont {Cotler}\ \emph {et~al.}(2020)\citenamefont {Cotler},
  \citenamefont {Jensen},\ and\ \citenamefont {Maloney}}]{cotler2020}%
  \BibitemOpen
  \bibfield  {author} {\bibinfo {author} {\bibfnamefont {J.}~\bibnamefont
  {Cotler}}, \bibinfo {author} {\bibfnamefont {K.}~\bibnamefont {Jensen}},\
  and\ \bibinfo {author} {\bibfnamefont {A.}~\bibnamefont {Maloney}},\
  }\bibfield  {title} {\bibinfo {title} {{Low-dimensional {de Sitter} quantum
  gravity}},\ }\href {https://doi.org/10.1007/jhep06(2020)048} {\bibfield
  {journal} {\bibinfo  {journal} {Journal of High Energy Physics}\ }\textbf
  {\bibinfo {volume} {2020}} (\bibinfo {year} {2020})}\BibitemShut {NoStop}%
\bibitem [{\citenamefont {Aalsma}\ and\ \citenamefont {Shiu}(2020)}]{shiu2020}%
  \BibitemOpen
  \bibfield  {author} {\bibinfo {author} {\bibfnamefont {L.}~\bibnamefont
  {Aalsma}}\ and\ \bibinfo {author} {\bibfnamefont {G.}~\bibnamefont {Shiu}},\
  }\bibfield  {title} {\bibinfo {title} {Chaos and complementarity in de
  {Sitter} space},\ }\bibfield  {journal} {\bibinfo  {journal} {Journal of High
  Energy Physics}\ }\textbf {\bibinfo {volume} {2020}},\ \href
  {https://doi.org/10.1007/jhep05(2020)152} {10.1007/jhep05(2020)152} (\bibinfo
  {year} {2020})\BibitemShut {NoStop}%
\bibitem [{\citenamefont {Gu}\ and\ \citenamefont {Kitaev}(2019)}]{Gu:2018jsv}%
  \BibitemOpen
  \bibfield  {author} {\bibinfo {author} {\bibfnamefont {Y.}~\bibnamefont
  {Gu}}\ and\ \bibinfo {author} {\bibfnamefont {A.}~\bibnamefont {Kitaev}},\
  }\bibfield  {title} {\bibinfo {title} {{On the relation between the magnitude
  and exponent of OTOCs}},\ }\bibfield  {journal} {\bibinfo  {journal} {Journal
  of High Energy Physics}\ }\textbf {\bibinfo {volume} {2019}},\ \href
  {https://doi.org/10.1007/jhep02(2019)075} {10.1007/jhep02(2019)075} (\bibinfo
  {year} {2019})\BibitemShut {NoStop}%
\end{thebibliography}%

\end{document}